\documentclass[manuscript,screen]{acmart}

\usepackage{booktabs}
\usepackage{multirow}
\usepackage{enumitem}
\usepackage{subfigure}
\usepackage[ruled]{algorithm2e}
\usepackage[flushleft]{threeparttable}
\usepackage{bbm}
\usepackage{mathrsfs}
\usepackage{longtable}
\usepackage{lscape} 

\usepackage{tikz}
\usepackage[edges]{forest}

\AtBeginDocument{%
  \providecommand\BibTeX{{%
    \normalfont B\kern-0.5em{\scshape i\kern-0.25em b}\kern-0.8em\TeX}}}

\usepackage{xcolor}
\usepackage{soul}
\usepackage{colortbl}

\newcommand{\eg}{\emph{e.g.}}
\newcommand{\ie}{\emph{i.e.}}

\definecolor{FeatEng}{rgb}{0.949,0.504,0.504}
\definecolor{FeatEco}{rgb}{0.582,0.879,0.824}
\definecolor{ScoreRank}{rgb}{0.836,0.965,0.676}
\definecolor{UserInter}{rgb}{0.996078,0.890196,0.54117}

\setcopyright{acmlicensed}
\copyrightyear{2018}
\acmYear{2018}
\acmDOI{XXXXXXX.XXXXXXX}

\acmConference[Conference acronym 'XX]{Make sure to enter the correct
  conference title from your rights confirmation emai}{June 03--05,
  2018}{Woodstock, NY}
\acmISBN{978-1-4503-XXXX-X/18/06}

\begin{document}

\title{A Survey on Diffusion Models for Recommender Systems}

\author{Jianghao Lin}
\affiliation{%
\institution{Shanghai Jiao Tong University}
\country{China}}
\email{chiangel@sjtu.edu.cn}

\author{Jiaqi Liu}
\affiliation{%
\institution{Shanghai Jiao Tong University}
\country{China}}
\email{1qaz2wsx3edc@sjtu.edu.cn}

\author{Jiachen Zhu}
\affiliation{%
\institution{Shanghai Jiao Tong University}
\country{China}}
\email{gebro13@sjtu.edu.cn}

\author{Yunjia Xi}
\affiliation{%
\institution{Shanghai Jiao Tong University}
\country{China}}
\email{xiyunjia@sjtu.edu.cn}

\author{Chengkai Liu}
\affiliation{%
\institution{Texas A\&M University}
\country{USA}}
\email{liuchengkai@tamu.edu}

\author{Yangtian Zhang}
\affiliation{%
\institution{Yale University}
\country{USA}}
\email{zytzrh@gmail.com}

\author{Yong Yu}
\affiliation{%
  \institution{Shanghai Jiao Tong University}
  \country{China}}
\email{yyu@sjtu.edu.cn}

\author{Weinan Zhang$^\dagger$}
\affiliation{%
  \institution{Shanghai Jiao Tong University}
  \country{China}}
\email{wnzhang@sjtu.edu.cn}

\renewcommand{\shortauthors}{J. Lin et al.}

\begin{abstract}

The rapid advancement of online services has positioned recommender systems (RSs) as crucial tools for mitigating information overload and delivering personalized content across e-commerce, entertainment, and social media platforms. 
While traditional recommendation techniques have made significant strides in the past decades, they still suffer from limited generalization performance caused by factors like inadequate collaborative signals, weak latent representations, and noisy data. 
In response, diffusion models (DMs) have emerged as promising solutions for recommender systems due to their distinguished generative capability, superior representation learning, and flexible internal structure.
To this end, in this paper, we present the first comprehensive survey on diffusion models for recommendation, and draw a bird’s-eye view from the perspective of the whole pipeline in real-world recommender systems. 
We systematically categorize existing research works into three primary domains: (1) diffusion for data engineering \& encoding that focuses on data augmentation and representation enhancement; (2) diffusion as recommendation models to directly estimate user preferences and rank items; and (3) diffusion for content presentation to generate personalized content such as fashion and advertisement creatives. 
Our taxonomy highlights the unique strengths of diffusion models in capturing complex data distributions and generating high-quality, diverse samples that closely align with user preferences. 
We also summarize the core characteristics of adapting diffusion models for recommendation, and further identify key areas for future exploration, which helps establish a roadmap for researchers and practitioners seeking to advance recommender systems through the innovative application of diffusion models. 
To further facilitate the research community of recommender systems based on diffusion models, we actively maintain a GitHub repository for papers and other related resources in this rising direction\footnote{\url{https://github.com/CHIANGEL/Awesome-Diffusion-for-RecSys}}.
\let\thefootnote\relax\footnotetext{$\dagger$ Weinan Zhang is the corresponding author.}
\end{abstract}



\received{20 February 2007}
\received[revised]{12 March 2009}
\received[accepted]{5 June 2009}

\maketitle

\section{Introduction}

With the rapid development of online services, recommender systems (RSs) have become increasingly indispensable to mitigate information overload problem~\cite{dai2021adversarial,fu2023f,liu2024mamba4rec} and match users’ information needs~\cite{guo2017deepfm,lin2023map}. 
They provide personalized suggestions across various scenarios such as movie~\cite{goyani2020review}, e-commerce~\cite{schafer2001commerce}, music~\cite{song2012survey}, etc. 
Despite the different forms of recommendation tasks (\eg, sequential recommendation, top-$N$ recommendation), the common objective for recommender systems is to precisely estimate a given user's preference towards each candidate item based on diverse source data (\eg, interaction data, user profile, item content), and finally arrange a ranked list of items presented to the user~\cite{lin2021graph,xi2023bird}. 

\begin{figure}[t]
    \centering
    \includegraphics[width=0.99\textwidth]{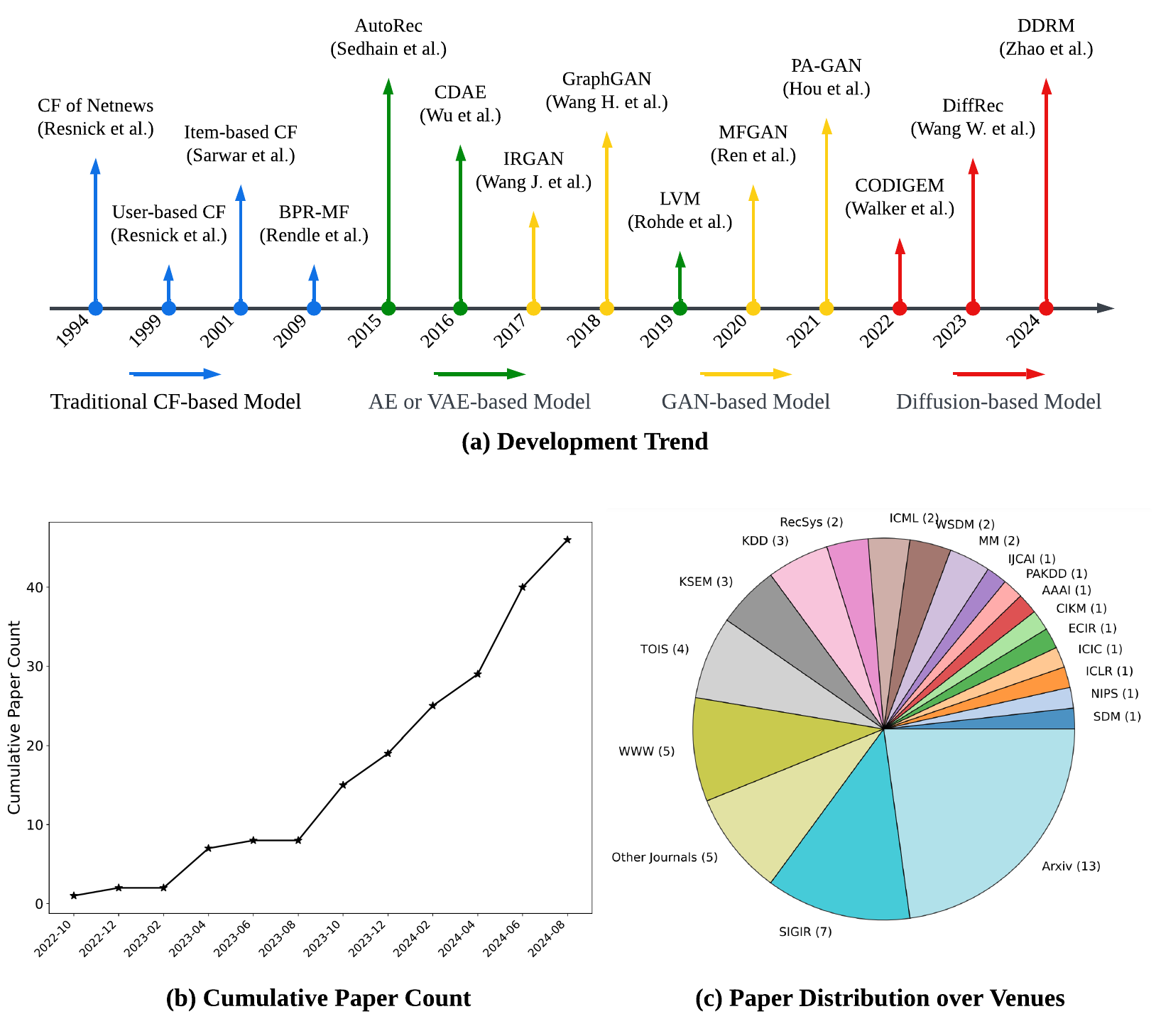}
    \caption{
    (a) The development trends and representational works of recommendation methods from traditional collaborative filtering (CF) based methods to generative methods, \ie, autoencoder (AE), variational autoencoder (VAE), generative adversarial network (GAN), and diffusion model.
    (b) The cumulative paper count of diffusion-based recommendation methods according to the timeline.
    (c) The paper distribution of diffusion-based recommendation methods over venues, where the venue name is followed by the exact number of published papers.
    }
    \label{fig:trend}
\end{figure}

As illustrated in Figure~\ref{fig:trend}, in the past decades, we have witnessed significant progress in the research on recommender systems, shifting from traditional techniques like collaborative filtering (CF)~\citep{he2016fast} to more advanced deep learning methodologies~\citep{lin2023mitigating}.
However, they usually suffer from limited generalization performance on account of inadequate collaborative signals~\citep{lin2023map}, weak latent representations~\citep{du2024disco}, noisy data scenarios~\citep{wang2021denoising}.
Therefore, the generative models, \eg, variational autoencoders (VAEs)~\citep{kingma2013auto_vae,rezende2014stochastic} and generative adversarial networks (GANs)~\citep{goodfellow_generative_2014,karras_style-based_2019,brock2018large_biggan}, turn out a promising solution to mitigate the above challenges for recommendation due to their generative nature and solid theoretical foundation. 
The self-supervised characteristic enables them to learn the non-linear probabilistic latent space and thereby precisely capture the users' dynamic and ever-evolving preferences~\cite{fan2022field,liang2024survey}.
However, these models still have their own limitations such as restricted representation capability~\citep{wang2023diffusion} and training instability~\citep{becker2022instability}. 

Recently, diffusion models (DMs)~\citep{ho2020denoising,song2020score} have emerged as
the state-of-the-art of generative modeling paradigms, achieving substantial success in various domains including computer vision~\citep{lugmayr2022repaint}, audio generation~\citep{lee2021nu}, natural language processing~\citep{austin2021structured}, and reinforcement learning~\citep{zhu2023diffusion}.
Unlike other earlier generative models such as VAEs and GANs, diffusion models leverage a denoising framework that effectively reverses a multi-step noising process to generate synthetic data that aligns closely with the distribution of the training data. 
This ensures the remarkable capabilities of diffusion models in capturing the multi-grained feature representations and generating high-quality samples, as well as maintaining enhanced training stability. 
We summarize the three main characteristics that make diffusion models attractive in the context of recommender systems:
\begin{enumerate}
    \item \textbf{Distinguished generative capability}. As one of the state-of-the-art generative paradigms, diffusion models are able to effectively capture the underlying distribution of the source data and fulfill various generative tasks to assist the downstream recommendation tasks, \eg, data imputation~\citep{zheng2022diffusion}, user behavior simulation~\citep{liu2023diffusion}, sample synthesis~\citep{yang2024balanced}, and image creation~\citep{yang2024new}.
    \item \textbf{Superior representation learning}. Diffusion models are powerful self-supervised methods, and are known for their remarkable abilities to learn high-quality, low-dimensional representations of the source data in a probabilistic generative manner~\citep{yang2024survey,fuest2024diffusion}. 
    In the context of recommender systems, they can effectively capture the underlying latent factors and representations based on multi-modal data (\eg, interaction data, complex user behaviors, and item attributes), thus making more accurate predictions about user preferences and generating recommendations tailored to individual tastes.
    \item \textbf{Flexible backbone structure}. As a general learning framework for generative modeling, the internal structures of diffusion models (\ie, the backbone design) are fairly flexible, and can be integrated with other deep learning models like U-Net~\citep{ho2020denoising} and Transformers~\citep{peebles2023scalable}.
    This allows for flexible backbone designs of diffusion models to effectively incorporate different types of heterogeneous information (\eg user demographics, temporal dynamics, and contextual cues) into the recommendation process.
\end{enumerate}

As a consequence, as shown in Figure~\ref{fig:trend}, an increasing range of pioneer attempts has been made to employ diffusion models for recommendation, achieving notable progress in boosting the performance of different canonical recommendation processes, \eg, data augmentation~\citep{wang2024diff}, user modeling~\citep{zhao2024denoising}, and content personalization~\citep{yang2024new}, etc. 
Given the advantages of applying diffusion models to recommendation, as well as the proliferation of research in the community, we believe that the time is right to conduct a comprehensive survey to systematically summarize the current research progress and provide inspirations for the future exploration, in terms of the adaption of diffusion models in recommender systems.

Diffusion models are closely related to generative modeling and self-supervised learning. 
There exist several related survey works that delve into the potential of generative models~\citep{li2023large,deldjoo2024review,li2024survey,xu2024survey,liang2024survey,deldjoo2021survey,zhang2020survey} or self-supervised learning techniques~\citep{jing2023contrastive,liu2023pre,yu2023self} for recommender systems. 
For example, \citet{liu2023pre} and \citet{yu2023self} conduct reviews on the self-supervised learning (SSL) based techniques for recommendation like contrastive learning and sequence modeling. 
\citet{deldjoo2021survey} centers on adversarial recommender systems, where generative adversarial networks (GANs) are widely employed for security and robustness.
\citet{liang2024survey} investigate the applications of variational autoencoders (VAEs) in recommender systems based on their generative Bayesian nature.
There is also a range of surveys~\citep{li2023large,deldjoo2024review,li2024survey,xu2024survey} focus on the generative recommendation with the assistance of large language models (LLMs), which serve as the most popular foundation models for various downstream scenarios in the past years.
However, none of these surveys concentrate on the applications of diffusion models for recommendation.
There still lacks a bird's-eye view of how recommender systems can embrace diffusion models and integrate them into different parts of the recommendation pipeline, which is essential in building a technical roadmap to systematically guide the research and industrial practice, of recommender systems empowered by diffusion models.

\begin{figure}[t]
    \centering
    \includegraphics[width=0.99\textwidth]{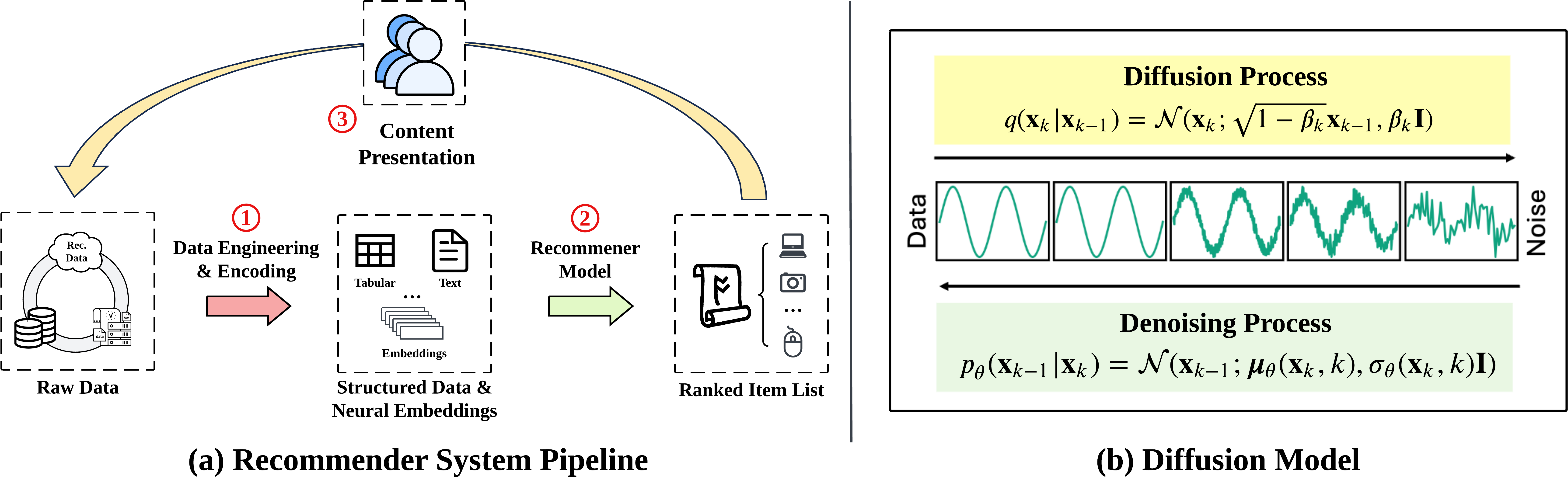}
    \caption{
    (a) The illustration of a deep learning based recommender system pipeline, which is characterized by three major stages: data engineering \& encoding, recommendation model, and content presentation. 
    (b) An overview of diffusion models for data analysis and generation via the diffusion-denoising process.
    }
    \label{fig:preliminary}
\end{figure}

To this end, in this paper, we aim to conduct a timely and comprehensive survey on the
adaption of diffusion models to recommender systems. 
As depicted in Figure~\ref{fig:preliminary}(a), we analyze the latest research progress and categorize the existing works according to different roles that diffusion models play in the modern recommender system pipeline:
\begin{itemize}
    \item \textbf{Diffusion for data engineering \& encoding}. Data engineering \& encoding is generally referred to as the process of manipulating and transforming the raw data collected online into structured data or neural embeddings for the downstream recommenders. 
    As a powerful class of generative models as well as self-supervised learning algorithms, diffusion models have shown remarkable capabilities in both \textit{data augmentation} and \textit{representation enhancement}, both of which help improve the downstream recommendation performance.
    \item \textbf{Diffusion as recommendation model}. The recommendation model aims to estimate a given user’s preference towards each candidate item, and finally arrange a ranked list of items presented to the user. 
    According to different types of tasks the recommender aims to solve, we classify the diffusion-based recommenders into three categories: \textit{collaborative recommendation}, \textit{context-aware recommendation}, and \textit{other applications}. 
    \item \textbf{Diffusion for content presentation}. Equipped with powerful generative capabilities of diffusion models, we can move one step further from personalized recommendation to individualized content generation. 
    That is, every single item can obtain different presentation contents (\eg, creatives, thumbnails) produced by diffusion models for different users or groups, which can largely promote the user satisfaction. 
    We categorize the research works in this line based on the different types of content to be generated: \textit{fashion generation}, \textit{ad creative generation}, and \textit{general content generation}.
\end{itemize}
Based on the taxonomy above, we can identify burgeoning trends within this rapidly evolving landscape, and therefore propose feasible and instructive suggestions for the evolution of existing online recommendation platforms considering the help of diffusion models.
The main contributions of this paper can be summarized as follows:
\begin{itemize}
    \item \textbf{Comprehensive and up-to-date review}. To the best of our knowledge, this is the first comprehensive, up-to-date and forward-looking survey on diffusion models for recommendation. 
    Our survey highlights the suitability of diffusion models for recommender systems and discusses the advantages they bring about from various aspects, varying from personalized recommendation to individualized content presentation. 
    \item \textbf{Unified and structured taxonomy}. We introduce a well-organized categorization to classify the existing research works into three major types according to the different roles the diffusion models play: diffusion for data engineering \& encoding, diffusion as recommendation model, and diffusion for content presentation.
    This taxonomy provides the readers with a coherent roadmap, and helps recognize the trend in applications of diffusion models to recommender systems from multiple perspectives.
    \item \textbf{Insights for challenges and future directions}. We highlight key challenges faced in the current research landscape, and further point out several promising directions for future exploration, aiming to shed light on recommender systems empowered by diffusion models and thereby attract more researchers to engage in this research field.
\end{itemize}

The remainder of this paper is organized as follows. 
In Section~\ref{sec:preliminary}, we briefly introduce the background and preliminary for
recommender systems and diffusion models.
In Section~\ref{sec:taxonomy}, we elaborate on the taxonomy of diffusion models for recommendation by categorizing  existing works into 
In Section~\ref{sec:challenge}, we highlight the limitations and
challenges shown in existing works, and discuss the potential future directions. 
Finally, we conclude the survey in
Section~\ref{sec:conclusion}.

\section{Preliminary}
\label{sec:preliminary}

Before elaborating on the details of our survey, we would like to introduce the following background and basic concepts: (1) the formulation and essential components of modern recommender systems, and (2) the general workflow and typical variants of diffusion models with the theoretical formula derivations.

\subsection{Modern Recommender Systems}
\label{sec:preliminary rs}

The core task of recommender systems is to arrange a ranked list of items $[i_k]_{k=1}^N, i_k \in \mathcal{I}$ for the user $u\in\mathcal{U}$ given a certain context $c\in\mathcal{C}$, where $\mathcal{U}$, $\mathcal{I}$ and $\mathcal{C}$ are the universal feature sets of users, items and contexts (\eg, device, time, season), respectively. 
Note that scenarios like next item prediction are special cases for such a formulation with $N=1$.
We formulate the goal of recommendation as follows:
\begin{equation}
    [i_k]_{k=1}^N\leftarrow \operatorname{RS}(u,c,\mathcal{I}),\; i_k\in\mathcal{I},u\in\mathcal{U},c\in\mathcal{C}.
\end{equation}
As shown in Figure~\ref{fig:preliminary}(a), there are generally three key components in deep learning based recommender system  pipelines. We briefly introduce each of them as follows:
\begin{itemize}
    \item \textbf{Data engineering \& encoding}. 
    The data pre-processing stage serves as the foundational pillar of modern deep learning based recommender systems, and generally consists of two primary processes: (1) data engineering and (2) data encoding. 
    Data engineering generally encompasses selecting, manipulating, transforming, and augmenting the raw data collected online into structured data that is suitable as inputs of neural recommendation models, which can reveal the important signals and patterns for user modeling.
    It does not only involve feature-level editing (\eg, cross-feature generation), but also includes sample-level synthesis (\eg, user interaction records generation and negative sample mining). 
    The structured outputs of data engineering possess various forms of features with different modalities, \eg, IDs, texts, images, audio, etc.
    The data encoding process takes as input the processed structured data from data engineering, and produces the corresponding neural embeddings for the downstream deep recommendation models. 
    Various encoders are employed depending on the data modality, \eg, vision models for visual features, language models for textual features.

    \item \textbf{Recommendation Model}. The deep recommendation model is the core algorithmic engine of a recommender system, tasked with selecting or ranking the top-relevant items to satisfy users’ information needs based on the structured data or neural embeddings produced by the data engineering \& encoding stage. 
    Researchers develop a variety of neural methods to accurately estimate the user interests and behavior patterns based on various techniques like sequence modeling~\citep{cheng2024general,liu2023user,liu2024behavior} and graph neural networks~\citep{wang2018ripplenet,wang2019neural}. 
    According to the task types and input data formats, the design of deep recommendation models can be generally classified into three categories: (1) collaborative recommendation, (2) context-aware recommendation, and (3) other applications.
    The collaborative recommendation is generally referred to as the collaborative filtering (CF) based methods simply based on the user-item co-occurrence matrix~\citep{he2017neural}. 
    The context-aware recommendation takes into account the contextual information surrounding a user's request to provide more precise recommendations, \eg, user behavior sequence for user profiling~\citep{zhou2018deep}, or knowledge graph for item content understanding~\citep{wang2018ripplenet}. 
    Finally, other applications refers to a broader range of tasks that are closely related to the aforementioned collaborative or context-aware recommendation, \eg, learning to rank~\citep{cao2007learning}, and bidding for online advertising~\citep{cai2017real}.
    \item \textbf{Content presentation}. After arranging the ranked list of items with the recommendation model, the content presentation serves as the final touch that brings the recommendations to life, ensuring the recommended items are delivered in a manner that is engaging, accessible, and contextually relevant to the target user. 
    Conventionally, it involves the strategic placement of recommended items, the manual usage of visual elements such as images, videos, and text to attract user attention, and the implementation of interactive features that encourage users to be more willing to explore and interact with the displayed items. 
    Despite some appealing manual designs, the content of one certain item presented to different users remains uniform.
    Nowadays, with the rapid development of generative models (\eg, large language models~\citep{zhao2023survey}, diffusion models~\citep{ho2020denoising}), we can take one step further from unified content presentation to individualized content generation. 
    That is, the concrete content of each recommended item (\eg, creatives, titles, thumbnails) can be dynamically edited or generated based on user profiles, device types, and environmental contexts, thereby increasing the likelihood of user attention and conversion. 
\end{itemize}

\subsection{Foundations of Diffusion Models}

Typically, as illustrated in Figure~\ref{fig:preliminary}(b), the training procedure of diffusion models includes two stages: the forward process (diffusion) and the reverse process (denoising).
In the forward diffusion process, the model turns a data sample into pure random noise by incrementally adding noise for multiple steps, which is usually a Markov process with each step depending only on the preceding one.
Then, the reverse denoising process learns to remove the noise to reconstruct the original data sample, essentially reversing the forward process. 
In this way, the model learns to remove the noise added during the diffusion process, and thereby generates samples from the same distribution as the training data.
According to different designs of the forward and backward processes, the common frameworks for diffusion models include denoising diffusion probabilistic models~\citep{ho2020denoising}, score-based generative methods~\citep{song_generative_2019, song2020score}, conditional diffusion models~\citep{rombach2022high}, etc. 
Next, we will provide a concise introduction to these three classical types of diffusion models.

\subsubsection{Denoising Diffusion Probabilistic Models (DDPMs)}
\label{sec:DDPM}

Denoising diffusion probabilistic models (DDPMs) are built upon a well-defined probabilistic process with dual Markov chains that consist of two parts: (1) the forward diffusion process that gradually transforms the data into pure noise with pre-determined noise (\eg, Gaussian noise), and (2) the reverse denoising process that aims to recover the original data via deep neural networks. 

\textbf{Forward Diffusion Process}. 
Assume that there is an initial clean data $\mathbf{x}_0 \sim q(\mathbf{x})$ 
sampled from a given data distribution $q(\mathbf{x})$. 
The ensuing forward diffusion process adulterates the initial data $\mathbf{x}_0$ by incrementally superimposing Gaussian noise, ultimately aiming to progress towards convergence with the standard Gaussian distribution (\ie, pure noise). 
During the forward process with maximum steps up to $K$, we will materialize a sequence of distributed latent data $[\mathbf{x}_1,\mathbf{x}_2,\cdots,\mathbf{x}_K]$, which can be formulated as a Markov chain transforming from $\mathbf{x}_{k-1}$ to $\mathbf{x}_k$ with a diffusion transition kernel:
\begin{equation} 
    q(\mathbf{x}_k | \mathbf{x}_{k-1}) = \mathcal{N}(\mathbf{x}_{k} ; \sqrt{1-\beta_{k}} \mathbf{x}_{k-1}, \beta_{k} \mathbf{I}), \;\forall {k} = 1, \dots, K ,
    \label{eq:ddpm-forward1}
\end{equation}
where $\beta_{k} \in (0,1)$ serves as a variance schedule to control the step size, $\mathbf{I}$ is the identity matrix with the same dimension as the input data $\mathbf{x}_{k-1}$, and $\mathcal{N}(\mathbf{x};\boldsymbol{\mu},\sigma\mathbf{I})$ is a Gaussian distribution of $\mathbf{x}$ with the mean $\mathbf{\mu}$ and the standard deviation $\sigma\mathbf{I}$. 
According to the property of the Gaussian kernel, we can get $\mathbf{x}_k$ directly from $\mathbf{x}_0$ by applying a series of transition kernels of Eq.~\ref{eq:ddpm-forward1}:
\begin{equation} 
\begin{aligned}
    q(\mathbf{x}_{k} | \mathbf{x}_{0}) & = \prod_{t=1}^k {q}(\mathbf{x}_{t} | \mathbf{x}_{t-1}) \\
    &= \mathcal{N}(\mathbf{x}_{k} ; \sqrt{\bar{\alpha}_{k}} \mathbf{x}_0, \sqrt{1 - \bar{\alpha}_{k}} \mathbf{I}),
    \label{eq:ddpm-forward2} 
\end{aligned}
\end{equation}
where $\alpha_{k} = 1 - \beta_{k}$, and $\bar{\alpha}_{k} = \prod_{i=1}^{k} \alpha_{i}$. 
Hence, we have:
\begin{equation}
    \mathbf{x}_{K} = \sqrt{\bar{\alpha}_{K}}  \mathbf{x}_0 + \sqrt{1 - \bar{\alpha}_{K}} \epsilon,
\end{equation}
where $\epsilon \sim \mathcal{N}(\mathbf{0},\mathbf{I})$ is the Gaussian noise. 
Specifically, it is designed $\bar{\alpha}_{K} \approx 0$ so that 
\begin{equation}
    q(\mathbf{x}_{K}) = \int q( \mathbf{x}_{K} | \mathbf{x}_{0} )q(\mathbf{x}_{0})\mathbf{d}\mathbf{x}_{0} \approx \mathcal{N}(\mathbf{x}_{K};\mathbf{0},\mathbf{I}).
    \label{eq:denoise appro}
\end{equation}
To sum up, the forward diffusion process gradually injects noise into the initial data $\mathbf{x}_0$ until it nearly aligns with the standard Gaussian distribution, and the reverse denoising process, as shown in Eq.~\ref{eq:denoise appro} can start with any Gaussian noise. 

\textbf{Reverse Denoising Process}. 
During the reverse denoising process, a series of Markov chain based transformations is employed until the original data $\mathbf{x}_0$ is reconstructed. 
To be specific, the series of reverse Markov chains should begin with a distribution $p(\mathbf{x}_K)=\mathcal{N}(\mathbf{x}_K;\mathbf{0},\mathbf{I})$.
Then, we maintain a learnable Gaussian transition kernel $p_\theta(\mathbf{x}_{k-1} | \mathbf{x}_k)$ to generate $p_\theta(\mathbf{x}_0)$:
\begin{equation} 
    {p}_\theta (\mathbf{x}_{k-1} | \mathbf{x}_{k}) = \mathcal{N}(\mathbf{x}_{k-1} ; \boldsymbol{\mu}_\theta(\mathbf{x}_{k}, {k}), \sigma_\theta(\mathbf{x}_{k}, {k}) \mathbf{I}), 
\label{eq:ddpm-reverse1}
\end{equation}
where the mean $\boldsymbol{\mu}_\theta(\cdot)$ and variance $\sigma_\theta(\cdot)$  are parameterized by $\theta$. 
The goal is to learn and approximate the data distribution via the model distribution ${p}_\theta(\mathbf{x}_0)$ during the reverse denoising process.

\textbf{Training}. 
In order to approximate the ground-truth data distribution $q(\mathbf{x})$, DDPM is trained to minimize the variational upper bound on the negative log-likelihood (NLL):

\begin{equation}
\begin{aligned}
\mathbb{E}\left[-\log p_{\theta}\left(\mathbf{x}_{0}\right)\right] &\leq \mathbb{E}_{q}\left[-\log \frac{p_{\theta}\left(\mathbf{x}_{0: K}\right)}{q\left(\mathbf{x}_{1: K} \mid \mathbf{x}_{0}\right)}\right] \\
& =\mathbb{E}_{q}\left[-\log p\left(\mathbf{x}_{K}\right)-\sum_{k \geq 1} \log \frac{p_{\theta}\left(\mathbf{x}_{k-1} \mid \mathbf{x}_{k}\right)}{q\left(\mathbf{x}_{k} \mid \mathbf{x}_{k-1}\right)}\right] \\
& =: L.
\end{aligned}
\label{eq:ddpm-training1}
\end{equation}
As suggested in \citep{ho2020denoising}, it can be rewritten using Kullback–Leibler divergence (KL divergence) as follows:
\begin{equation} \label{eq:ddpm-training2}
\begin{aligned}
L = \mathbb{E}_{q}\big[\underbrace{D_{\mathrm{KL}}\left(q\left(\mathbf{x}_{K} \mid \mathbf{x}_{0}\right) \| p\left(\mathbf{x}_{K}\right)\right)}_{L_{K}}  +\sum_{k>1} \underbrace{D_{\mathrm{KL}}\left(q\left(\mathbf{x}_{k-1} \mid \mathbf{x}_{k}, \mathbf{x}_{0}\right) \| p_{\theta}\left(\mathbf{x}_{k-1} \mid \mathbf{x}_{k}\right)\right)}_{L_{k-1}} \underbrace{-\log p_{\theta}\left(\mathbf{x}_{0} \mid \mathbf{x}_{1}\right)}_{L_{0}}\big],
\end{aligned}
\end{equation}
where the loss is decomposed into three parts: the prior loss $L_K$, the divergence of the forwarding step and the corresponding reversing step $L_{k-1}$, and the reconstruction loss $L_0$.
We can maximize the log-likelihood by only training the divergence loss between two steps $L_{k-1}$, and parameterize the posterior $q\left(\mathbf{x}_{k-1} \mid \mathbf{x}_{k}, \mathbf{x}_{0}\right)$ based on:
\begin{equation}
\begin{aligned}
    q\left(\mathbf{x}_{k-1} \mid \mathbf{x}_{k}, \mathbf{x}_{0}\right) &=\mathcal{N}\left(\mathbf{x}_{k-1} ; \tilde{\boldsymbol{\mu}}_{k}\left(\mathbf{x}_{k}, \mathbf{x}_{0}\right), \tilde{\beta}_{k} \textbf{I}\right), \\
    \tilde{\boldsymbol{\mu}}_k(\mathbf{x}_k, \mathbf{x}_0) &= \frac{\sqrt{\bar\alpha_{k-1}}\beta_k }{1-\bar\alpha_k}\mathbf{x}_0 + \frac{\sqrt{\alpha_k}(1- \bar\alpha_{k-1})}{1-\bar\alpha_k} \mathbf{x}_k \hspace*{1mm}, \\
    \tilde\beta_k &= \frac{1-\bar\alpha_{k-1}}{1-\bar\alpha_k}\beta_k,
\end{aligned}
\end{equation}
where $\alpha_k = 1-\beta_k$ and $\bar{\alpha}_k=\prod_{i=1}^{k} \alpha_{i}$. 
In this way, $L_{k-1}$ can be equally regarded as the expected value of the L2 loss between the two mean coefficients:
\begin{equation}
L_{k-1}=\mathbb{E}_{q}\left[\frac{1}{2 \sigma_{k}^{2}}\left\|\tilde{\boldsymbol{\mu}}_{k}\left(\mathbf{x}_{k}, \mathbf{x}_{0}\right)-\boldsymbol{\mu}_{\theta}\left(\mathbf{x}_{k}, k\right)\right\|^{2}_2\right]+C.
\end{equation}
Moreover, rather than predicting the mean $\boldsymbol{\mu}_\theta(\mathbf{x}_k,k)$, we can estimate the noise vector (matrix) to be eliminated at each time step by parameterizing $\epsilon_\theta(\mathbf{x}_k,k)$ for simplification:
\begin{equation}
    \mathbb{E}_{k \sim \mathcal{U} (1,K), \mathbf x_0 \sim q(\mathbf x), \epsilon \sim \mathcal{N}(\mathbf{0},\mathbf{I})} \bigg[ {\lambda(k)  \left\| \epsilon - \epsilon_\theta(\mathbf{x}_k, k) \right\|^2_2} \bigg],
\end{equation}
where $\lambda(k) = \frac{{\beta_k}^2}{2{\sigma_K}^2\alpha_k(1-\bar{\alpha}_k)}$ 
is the weight coefficient for noise scale, and $\epsilon_\theta(\mathbf{x}_k, k)$ is a model for step-wise Gaussian noise prediction. 
The model $\epsilon_\theta$, trained based on the loss function above, will then be employed for the sampling inference of the reverse denoising process. 

\textbf{Inference (Sampling)}. 
Given the noisy data $\mathbf{x}_K$, we start the $K$-step reverse denoising process and gradually generate the data $\mathbf{x}_0$ as follows:
\begin{equation}
\begin{aligned}
    p_\theta(\mathbf{x}_{k-1}|\mathbf{x}_k) &= \mathcal{N}(\mathbf{x}_{k-1}; \boldsymbol{\mu}_\theta(\mathbf{x}_k, k), \sigma_\theta(\mathbf{x}_k, k)\mathbf{I})  \\ 
    &= \frac{1}{\sqrt{\alpha_k}}(\mathbf{x}_k - \frac{\beta_k}{\sqrt{1-\overline{\alpha_k}}}\epsilon_\theta(\mathbf{x}_k, k)) + \sigma_\theta(\mathbf{x}_k, k)z,
\end{aligned}
\end{equation}
where $z \sim \mathcal{N}(\mathbf{0}, \mathbf{I})$, and $\beta_k \approx \sigma_\theta^2(\mathbf{x}_k, k)$.
While the vanilla DDPM~\citep{ho2020denoising} are proposed to handle continuous data like audio~\citep{lee2021nu} and image~\citep{lugmayr2022repaint}, researchers have extended its applications to other data mining scenarios like tabular data~\citep{kotelnikov2023tabddpm} and textual data~\citep{zhu2023diffusion} with specially designed adaptations.

\subsubsection{Score-based Generative Models (SGMs)} 
\label{sec:SDE}
Score-based generative models (SGMs)~\citep{song_generative_2019, song2020score} further generalize DDPM's discrete diffusion processes to a continuous framework based on stochastic differential equations (SDEs)~\citep{van1976stochastic}. 
For clarity, here we adopt the notation $t\in[0,T]$ for SGMs instead of the step size $k=1,\dots,K$ in DDPMs. 
Consequently, the sequence $\mathbf{x}_0,\dots,\mathbf{x}_K$ is replaced
with a continuous function $\mathbf{x}(t)$. 
Note that here we discuss the continuous property of the diffusion-denoising process, while the data itself can be either continuous (\eg, images) or discrete (\eg, texts).

\textbf{Forward Diffusion Process.} 
The continuous diffusion process can be formulated based on SDEs, consisting of a mean shift and a Brownian motion (\ie, standard Wiener process) as follows:
\begin{equation}
	\mathrm{d} \mathbf{x}=\mathbf{f}(\mathbf{x}, t) \mathrm{d} t+g(t) \mathrm{d} \mathbf{w}, \; t \in [0, T],
\end{equation}
where $\mathbf{f}(\cdot, t)$ denotes the drift coefficient for the stochastic continuous process $\mathbf{x}(t)$, and $g(\cdot)$ is the diffusion coefficient interwined with the Brownian motion $\mathbf{w}$.

Similar to DDPMs, $\mathbf{x}_0$ and $\mathbf{x}_T$ are sampled from the clean distribution $p_0=\mathcal{N}(\mathbf{x}_0;\mathbf{0},\mathbf{I})$ 
and the standard Gaussian distribution $p_T=\mathcal{N}(\mathbf{x}_T;\mathbf{0},\mathbf{I})$, respectively. The generalized continuous version of DDPM, also known as Variance Preserving SDE (VP-SDE), can be written as:
\begin{equation}
    \mathrm{d} \mathbf{x}=-\frac{1}{2} \beta(t) \mathbf{x} \mathrm{d}t+\sqrt{\beta(t)} \mathrm{d} \mathbf{w}. 
\end{equation}

\textbf{Reverse Denoising Process.} 
We can synthesize the new sample from the known prior distribution $p_T$ by solving the reverse-time SDE:
\begin{equation}
	\mathrm{d} \mathbf{x}=\left[\mathbf{f}(\mathbf{x}, t)-g^2(t) \nabla_\mathbf{x} \log p_t(\mathbf{x}) \right] \mathrm{d} t+g(t) \mathrm{d} \bar{\mathbf{w}},
\end{equation}
where $\bar{\mathbf{w}}$ is the reverse Brownian motion~\citep{vincent2011connection}, $p_t(\mathbf{x})$ is the probability density of $\mathbf{x}(t)$, and $s(\mathbf{x})=\nabla_\mathbf{x} \log p_t(\mathbf{x})$ is called the score function of $p_t(\mathbf{x})$. 
In practice, we would maintain a parameterized time-dependent neural network $s_\theta(\mathbf{x}, t)$ to estimate the score function, which can be optimized by minimizing:
\begin{equation}
L =  \mathbb{E}_{t,\mathbf{x}_0,\mathbf{x}_t} \left[\lambda(t) \|{s}_{\boldsymbol{\theta}}(\mathbf{x}_t, t)-\nabla_{\mathbf{x}_t} \log p(\mathbf{x}_t| \mathbf{x}_0)\|_{2}^{2}\right], 
\end{equation}
where $\lambda(t)$ is the weighting function, and $\mathbf{x}_0$ is sampled from the clean distribution $p_0$. 
In this way, we avoid the direct estimation of the impractical score function by calculating the transition probability which adheres to a Gaussian distribution throughout the forward diffusion process~\citep{song2020score}.

Upon finishing the training process, we can generate samples based on various techniques like the Euler-Maruyama (EM)~\citep{mao2015truncated}, Prediction-Correction (PC)~\citep{butcher2016numerical}, or Probability Flow ODE method~\citep{song2020score}.

\subsubsection{Conditional Diffusion Models}
In the previous sections, we have introduced the DDPMs and SGMs from an unconditional perspective, where they generate the data samples based on the learned distribution of the source data without any explicit guidance or conditions. 
However, the ability to control the generation process by passing explicit guidance or conditions is an important characteristic of generative models.
The diffusion models are able to generate the data samples not only from an unconditional distribution $p_0$, but also from a conditional distribution $p_0(\mathbf{x}|c)$ given a condition $c$. 
The conditioning signals can have a variety of modalities, ranging from class labels to features (\eg, text embeddings) related to the input data $\mathbf{x}$~\citep{rombach2022high}. 
More specifically, there are various sampling algorithms designed for conditional generation~\citep{yang2023diffusion}, \eg, label-based guidance~\citep{dhariwal2021diffusion}, label-free guidance~\citep{ho2022classifier}, text-based conditions~\citep{le2024voicebox,gong2022diffuseq}, graph-based conditions~\citep{schneuing2022structure}, etc. 

Classically, the sampling under the conditions of labels and classifiers involves using gradient guidance at each step, typically requiring an additional differential classifier $p_{\phi}(c | \mathbf{x})$ (\eg, U-Net~\citep{ronneberger2015u} and Transformer~\citep{vaswani2017attention}) to generate condition gradients for specific labels~\citep{dhariwal2021diffusion}. 
These guidance labels are flexible, and can be textual, categorical, or task-specific feature embeddings~\citep{dhariwal2021diffusion,nichol2021glide,hu2022global,yang2024generate}. 
This is referred to as the classifier guidance, whose conditional reverse process can be written as:
\begin{equation}
    p_{\theta,\phi}(\mathbf{x}_{k-1}\mid \mathbf{x}_k,c) = Z p_{\theta}(\mathbf{x}_{k-1}\mid \mathbf{x}_k)p_{\phi}(c\mid\mathbf{x}_{k-1}),
\end{equation}
where $Z$ is the normalization factor. 

Although classifier guidance is a common and versatile approach to improve the sample quality, it heavily relies on the availability of a noise-robust pre-trained classifier $p_{\phi}(c|\mathbf{x})$. 
This requirement largely depends on the existence of annotated data to well train the classifier network, which is impractical in many real-world data-hungry applications. 
To this end, the classifier-free guidance is proposed.
Compared to the high accuracy of the labeled conditional diffusion model, the sampling under unlabeled conditions solely relies on self-information for guidance, and is better at generating innovative, creative and diverse data samples~\citep{choi2021ilvr,epstein2023diffusion,chao2022investigating,kollovieh2024predict}. 
The classifier-free guidance is usually the dense representations produced by specific encoder models based on certain contextual information like textual data, sequential data and graphic data.

As a result, both the classifier-based and classifier-free conditional diffusion models have been widely adopted in various online application services due to their high-quality and well-controlled generative outputs~\citep{bansal2023universal,zhang2023adding,kollovieh2024predict,nichol2021glide}.

\section{Diffusion Models for Recommendation: Taxonomy}
\label{sec:taxonomy}

Based on the decomposition of modern recommender systems discussed in Section~\ref{sec:preliminary rs}, we introduce the taxonomy framework of diffusion models for recommendation according to different roles that diffusion models play at different parts of the modern deep learning based recommender system pipeline: (1) diffusion for data engineering \& encoding, (2) diffusion as recommendation model, and (3) diffusion for content presentation. 
We depict the overall taxonomy framework in Figure~\ref{fig:taxonomy}, and list the research works of each category to facilitate readers in quickly indexing and finding the corresponding works.

\tikzstyle{level_zero}=[
    text centered,
    align=center,
    fill={rgb,255:red,242; green,243; blue,245}
]

\tikzstyle{level_one}=[
    text centered,
    align=center,
    text width=10em,
    fill={rgb,255:red,255; green,254; blue,179}
]

\tikzstyle{level_two}=[
    text centered,
    align=center,
    text width=13em,
    fill={rgb,255:red,233; green,247; blue,227}
]

\tikzstyle{level_three}=[
    text width=11.4em,
    align=left
]

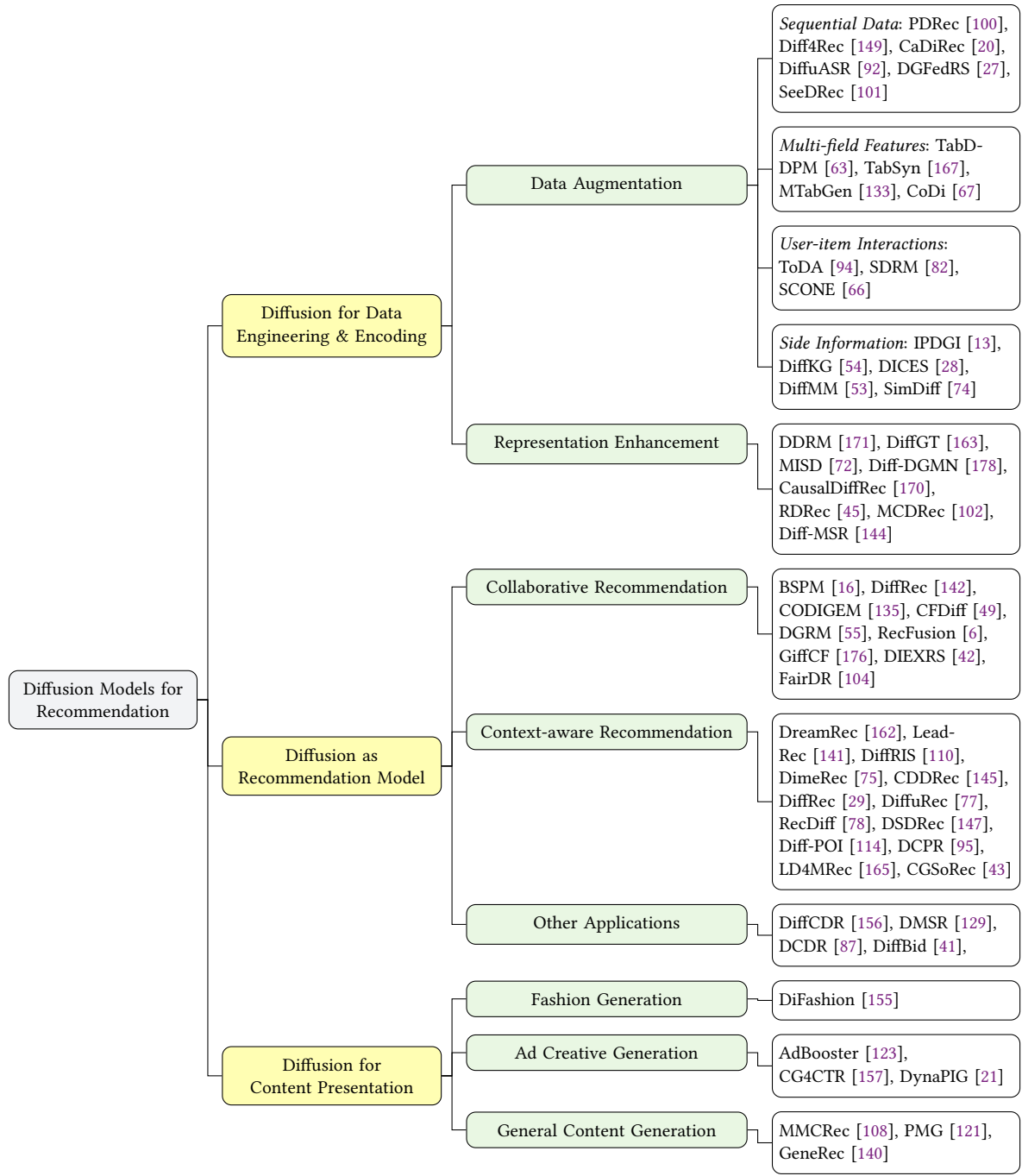
\begin{figure*}[tp]
  \centering
  \begin{forest}
    forked edges,
    for tree={
      grow=east,
      reversed=true,
      anchor=base west,
      parent anchor=east,
      child anchor=west,
      base=left,
      font=\small,
      rectangle,
      draw=black,
      rounded corners,
      minimum width=2.5em,
      s sep=6pt,
      inner xsep=3pt,
      inner ysep=5pt,
      ver/.style={rotate=90, child anchor=north, parent anchor=south, anchor=center},
    },
    where level=1{text width=5.6em,font=\small}{},
    where level=2{text width=5.3em,font=\small}{},
    where level=3{text width=9.7em,font=\small}{},
    [Diffusion Models for \\ Recommendation, level_zero, text width=8.5em
        [Diffusion for Data \\ Engineering \& Encoding, level_one
            [Data Augmentation,level_two
               [\textit{Sequential Data}: PDRec~\cite{ma2024plug}{,} Diff4Rec~\cite{wu2023diff4rec}{,} CaDiRec~\cite{cui2024diffusion}{,} DiffuASR~\cite{liu2023diffusion}{,} DGFedRS~\cite{difederated}{,} SeeDRec~\cite{ma2024seedrec},level_three]
               [\textit{Multi-field Features}: TabDDPM~\cite{kotelnikov2023tabddpm}{,} TabSyn~\cite{zhang2023mixed}{,} MTabGen~\cite{villaizan2024diffusion}{,} CoDi~\cite{lee2023codi},level_three]
               [\textit{User-item Interactions}: ToDA~\cite{liu2024toda}{,} SDRM~\cite{lilienthal2024multi}{,} SCONE~\cite{lee2024stochastic},level_three]
               [\textit{Side Information}: IPDGI~\cite{chen2023adversarial}{,} DiffKG~\cite{jiang2024diffkg}{,} DICES~\cite{dong2024dices}{,} DiffMM~\cite{jiang2024diffmm}{,} SimDiff~\cite{li2024simdiff},level_three]
            ]
            [Representation Enhancement,level_two
                [DDRM~\cite{zhao2024denoising}{,} DiffGT~\cite{yi2024directional}{,} MISD~\cite{li2024multi}{,} Diff-DGMN~\cite{zuo2024diff}{,} CausalDiffRec~\cite{zhao2024graph}{,} RDRec~\cite{he2024diffusion}{,} MCDRec~\cite{ma2024multimodal}{,} Diff-MSR~\cite{wang2024diff},level_three]
            ]
        ]
        [Diffusion as \\ Recommendation Model, level_one
            [Collaborative Recommendation,level_two
               [BSPM~\cite{choi2023blurring}{,} DiffRec~\cite{wang2023diffusion}{,} CODIGEM~\cite{walker2022recommendation}{,} CFDiff~\cite{hou2024collaborative}{,} DGRM~\cite{jiangzhou2024dgrm}{,} RecFusion~\cite{benedict2023recfusion}{,} GiffCF~\cite{zhu2024graph}{,} DIEXRS~\cite{guo2023explainable}{,} FairDR~\cite{malitesta2024fair},level_three]
            ]
            [Context-aware Recommendation,level_two
                [DreamRec~\cite{yang2024generate}{,} LeadRec~\cite{wang2024leadrec}{,} DiffRIS~\cite{niu2024diffusion}{,} DimeRec~\cite{li2024dimerec}{,} CDDRec~\cite{wang2024conditional}{,} DiffRec~\cite{du2023sequential}{,} DiffuRec~\cite{li2023diffurec}{,} RecDiff~\cite{li2024recdiff}{,} DSDRec~\cite{wang2024dsdrec}{,} Diff-POI~\cite{qin2023diffusion}{,} DCPR~\cite{long2024diffusion}{,} LD4MRec~\cite{yu2023ld4mrec}{,} CGSoRec~\cite{he2024balancing},level_three]
            ]
            [Other Applications,level_two
                [DiffCDR~\cite{xuan2024diffusion}{,} DMSR~\cite{tomasi2024diffusion}{,} DCDR~\cite{lin2024discrete}{,} DiffBid~\cite{guo2024aigb}{,} ,level_three]
            ]
        ]
       [Diffusion for \\ Content Presentation, level_one
            [Fashion Generation,level_two
               [DiFashion~\cite{xu2024diffusion},level_three]
            ]
            [Ad Creative Generation,level_two
                [AdBooster~\cite{shilova2023adbooster}{,} CG4CTR~\cite{yang2024new}{,} DynaPIG~\cite{czapp2024dynamic},level_three]
            ]
            [General Content Generation,level_two
                [MMCRec~\cite{mukande2024mmcrec}{,} PMG~\cite{shen2024pmg}{,} GeneRec~\cite{wang2023generative},level_three]
            ]
       ]
    ]
    ]
  \end{forest}
  \caption{The overall categorization of diffusion models for recommendation. We also list the research works of each category attached with the corresponding method name and reference. 
  }
  \label{fig:taxonomy}
\end{figure*}


\subsection{Diffusion for Data Engineering \& Encoding}
\label{sec:diffusion for data engineering and encoding}

Data engineering and encoding encompass the sophisticated processes of refining and converting the vast array of raw data gathered from online sources into structured formats or advanced neural embeddings, tailored for the efficient functioning of downstream recommendation systems. 
While the current recommender systems generally suffer from problems like data sparsity~\cite{wu2023diff4rec}, noisy data~\cite{zhao2024denoising} and filtering bubble~\cite{piao2023human}, diffusion models emerge as an exceptionally potent category of generative models to mitigate such issues during the data engineering and encoding stage.
They exhibit outstanding prowess in the dual realms of \textit{data augmentation} and \textit{representation enhancement}, which are pivotal in significantly bolstering the overall effectiveness and precision of downstream recommendation performance. 
Diffusion for data augmentation refers to the process of manipulating and augmenting the original raw training data (\eg, ID features, user behavior sequences, and user-item interactions) in data engineering phase, while diffusion for representation enhancement primarily focuses on strengthening and enhancing the neural embeddings (\eg, user \& item embeddings, and multi-modal representations) during the data encoding phase for downstream recommendation models. 

\subsubsection{Diffusion for Data Augmentation}
\label{sec:diffusion for data augmentation}

As a powerful class of generative models, diffusion models are able to capture the underlying distribution of given data and synthesize realistic samples for the downstream recommenders. 
By augmenting the original training data, diffusion models can thereby improve the recommender systems from various perspectives, \eg, handling sparsity, enhancing diversity, reducing noise, and improving model generalization. 
In the following, we will discuss leveraging diffusion models for data augmentation according to different data types to be generated for augmentation, \ie, \textit{sequential data augmentation}, \textit{feature imputation}, \textit{user-item interaction synthesis}, and \textit{side information editing}. 

First and foremost, \textbf{\textit{sequential data}} (\eg, user behavior sequence) serves as one of the core data types for recommendation to capture dynamic user preferences by modeling the sequential dependencies among historical user behaviors. 
Diffusion models can augment the sequence by either inserting, replacing, or reweighting user behaviors at item level~\cite{wu2023diff4rec,cui2024diffusion,ma2024plug}, or directly synthesize the whole sequence from scratch under certain guidance at sequence level~\cite{liu2023diffusion,difederated}. 
Diff4Rec~\cite{wu2023diff4rec} employs a curriculum-scheduled diffusion augmentation
framework for sequential recommendation by corrupting and reconstructing the user-item interactions (\ie, behaviors) in the latent space, and the generated outputs are progressively fed into the sequential recommenders with an easy-to-hard scheduler.
CaDiRec~\cite{cui2024diffusion} designs a context-aware diffusion-based contrastive learning method for sequential recommendation.
Given a user sequence, the model selects certain positions and generates alternative items for these positions with the guidance of context information (\ie, the sequential dependencies), which creates semantically consistent augmented views of the original sequence for contrastive learning.
DGFedRS~\cite{difederated} concentrate on the data sparsity issues in sequential federated recommender systems by capturing diverse latent user preferences and suppressing noise. 
It partitions the user behavior history and extends each segment into a longer synthetic behavior sequence via a guided diffusion generation, where a step-wise scheduling strategy is designed to control the data noise.
DiffuASR~\cite{liu2023diffusion} proposes a diffusion-based pseudo sequence generation framework, and fills in the gap between the generations of continuous images and discrete sequences by designing a sequential U-Net under two guided conditions (\ie, classifier-guided condition, and classifier-free condition). 
SeeDRec~\cite{ma2024seedrec} utilizes diffusion models to reformulate the user interest distribution from item level to sememe level to make full use of the deep semantic knowledge, which acts as the prompts to facilitate the downstream sequential recommendation.

Apart from the sequential data, \textbf{\textit{multi-field categorical/numerical features}} (or say tabular data) are also of great importance for recommender systems to capture the dynamic user preferences by modeling feature crossings~\cite{lin2023map,guo2017deepfm,wang2017deep}.
Hence, diffusion models are widely utilized for feature generation and imputation~\cite{villaizan2024diffusion,kotelnikov2023tabddpm,zhang2023mixed,lee2023codi}. 
For example, TabDDPM~\citep{kotelnikov2023tabddpm} introduces the design of DDPM~\cite{ho2020denoising} to the field of tabular data, and enables it to synthesizing a mixed types of data including ordinal, numerical, and categorical features. 
TabSyn~\cite{zhang2023mixed} involves the diffusion model within a variational autoencoder (VAE) crafted latent space, and achieves high-quality for mixed-type feature generation with faster generation speed via a linear noise schedule (\ie, less than 20 reverse steps). 
\citet{villaizan2024diffusion} further incorporate a full encoder-decoder transformer with diffusion process as the denoising model, which allows for the
conditioning attention mechanism while effectively capturing and representing complex interactions and dependencies among the input features.

Other works leverage diffusion models to simulate the \textbf{\textit{user-item interactions}} as the augmented training data for recommendation models with various purposes, \eg, cold-start problems~\cite{wu2023diff4rec}, hard negative sample mining~\citep{lee2024stochastic}, privacy-preserving concerns~\cite{lilienthal2024multi}, and adversarial robustness~\cite{liu2024toda}. 
For instance, \citet{liu2024toda} propose a target-oriented diffusion attack model to generate deceptive user interactions profiles with guidance of cross-attention mechanisms for shilling attacks towards the target item, which helps deepen the insights of the vulnerabilities and adversarial attacks in recommender systems.
SDRM~\cite{lilienthal2024multi} employs diffusion models to capture complex patterns of real-world datasets, and generates high-quality synthetic user-item interactions to augment or even replace the original dataset, aiming at addressing the data sparsity problem, as well as the privacy and security concerns for recommendation. 

Finally, diffusion models are also widely used to edit and augment the \textbf{\textit{side information}} for recommender systems, including but not limited to multi-modal data~\cite{chen2023adversarial,yang2024new,jiang2024diffmm} and knowledge graphs~\cite{dong2024dices,jiang2024diffkg,jiang2024diffmm,li2024simdiff}.
For example, \citet{chen2023adversarial} investigate the vulnerability of visually-aware recommender systems, and utilize a guided diffusion model to generate high-fidelity adversarial images designed to promote the exposure rates of specific items. 
They incorporate a conditional constraint into the diffusion process to ensure the  edited images closely resemble the original ones, and generate imperceptible perturbations to align the visual features of target items with popular items. 
DiffKG~\cite{jiang2024diffkg} concentrate on eliminating the irrelevant and noisy relations in knowledge graphs for recommender systems. It introduces a collaborative knowledge graph convolution mechanism that uses collaborative signals to guide the diffusion process for graph structure denoising, thereby aligning item semantics with collaborative relation modeling and leading to precise recommendations.

\subsubsection{Diffusion for Representation Enhancement}

Different from directly augmenting the raw input data during the data engineering phase as introduced above, diffusion for representation enhancement generally focuses on capturing the underlying distribution of raw input data and transforming it into robust feature embeddings to assist the training of downstream recommendation models.
As a unique instance of self-supervised learning paradigms with explicit denoising process, diffusion models are capable of establishing generalized latent spaces for enhanced representations and therefore addressing several key challenges of recommender systems, \eg, the multi-interest and ever-evolving user preference~\cite{liu2024mamba4rec,xi2024memocrs}, the diverse item aspects~\cite{fan2021modeling,fan2022sequential}, and the noise and uncertainty of interaction data~\cite{hurley2011novelty,wang2021clicks}.
A range of related studies~\cite{zhao2024denoising,yi2024directional,he2024diffusion,zhao2024graph,ma2024multimodal,wang2024diff,zuo2024diff,li2024multi} have been proposed in this line of research, each of which incorporates diffusion models for enhanced representation learning under different goals and scenarios. 
We briefly introduce these research works as follows.

DDRM~\cite{zhao2024denoising}, DiffGT~\cite{yi2024directional} and MISD~\cite{li2024multi} aim to \textbf{\textit{denoise the implicit user feedback}}. 
DDRM~\cite{zhao2024denoising} attempts to robustify the user and item representations for arbitrary recommendation models. It injects controlled Gaussian noise into user and item embeddings during a forward phase and then iteratively removes this noise during a reverse denoising phase, guided by a specialized denoising module utilizing the collaborative signals. 
Besides, in the inference stage, DDRM leverages the average embeddings of the user's historically interacted items as the starting point rather than a pure noise vector to further promote the personalization. 
DiffGT~\cite{yi2024directional} further considers handling the noisy implicit user feedback for neural graph recommenders. 
The authors showcase the anisotropic nature of recommendation data and propose to incorporates anisotropic directional Gaussian noise to improve the diffusion process, ensuring that the forward noise better aligns with the observed characteristics of recommendation data.
The model also integrates a graph transformer architecture with a linear attention module to efficiently robustify the noisy embeddings, which is guided by personalized information to improve the estimation of user preferences. 
MISD~\cite{li2024multi}, on the other hand, leverages diffusion models to address the noisy feedback in multi-interest user modeling, targeting multi-behavior sequential recommendation.

CausalDiffRec~\cite{zhao2024graph} employs causal diffusion models to \textbf{\textit{address the out-of-distribution (OOD) issue}} in the field of graph-enhanced recommendation. 
It incorporates backdoor adjustment and variational inference to capture the real environmental distribution, and then uses it as prior knowledge to guide the reverse phase of the diffusion process for invariant representation learning, which eliminates the impact of environmental confounders. 
The authors also provide theoretical derivations to prove that the proposed objective of CausalDiffRec encourages the model to learn environment-invariant graph representations, achieving excellent generalization performance in recommendations under distribution shifts and OOD data. 

RDRec~\cite{he2024diffusion} focuses on \textbf{\textit{improving the review-based textual embeddings}} to better model the diverse and dynamic user interests when facing different items. 
The model corrupts the user representations by adding noises to the original review-based textual features of the user interaction sequence, and the perturbed user representations undergo denoising via a transformer approximator with the awareness of target item information. 
Consequently, the model is able to learn to capture the dynamic user preferences and therefore generate generalized user interest embeddings for final recommendation.

MCDRec~\cite{ma2024multimodal} tailors diffusion models to \textbf{\textit{fuse the multi-modal knowledge}} for representation learning in multi-modal recommendation. 
Specifically, it incorporates the pre-extracted multi-modal information as conditions for the diffusion training process, aiming to fuse the conditional multi-modal knowledge into the generation of item representations. 
Moreover, the multi-modal diffused representations can be further utilized to denoise and reconstruct the user-item bipartite graph by computing the diffusion-aware interaction probability and filtering the occasional interactions. 
In this way, diffusion models serve as the core bridge to mitigate the bias between the multi-modal features and collaborative signals, and thereby enhance the item representations for improved recommendation performance. 

Diff-MSR~\cite{wang2024diff} concentrates on the \textbf{\textit{cold-start problem in multi-scenario recommendation}}, and leverages the transfer capabilities of diffusion models to enhance the representation learning of long-tail cold-start scenarios, utilizing information from other scenarios with ample training data.
Built upon a pretrained multi-scenario recommendation model, the authors design a piece-wise variance schedule, and then train a cold-or-rich domain classifier to obtain the candidates from rich domains (if incorrectly classified) to generate high-quality and informative embeddings for cold-start domains. 
It turns out that diffusion models are capable of capturing the commonality and distinction of various scenarios, enabling effective knowledge transferring among cold-start and rich domains. 

\begin{figure}[t]
    \centering
    \includegraphics[width=0.99\textwidth]{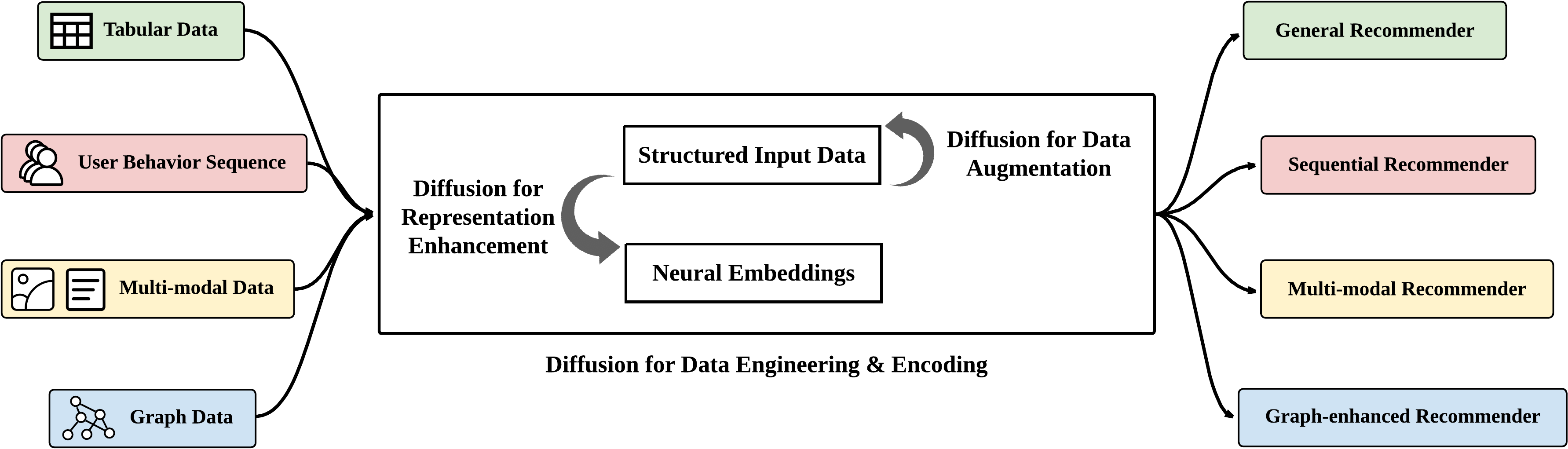}
    \caption{The illustration of the core characteristic (\ie, flexibility) of diffusion models when adapting them to the data engineering \& encoding stage in recommender systems. Diffusion models are generally compatible with various upstream input data types and downstream recommendation models. 
    }
    \label{fig:flexibility property}
\end{figure}

\subsubsection{Discussion}
\label{sec:discussion of diffusion for data engineering and encoding}

When adapting diffusion models for the data engineering \& encoding stage, the diffusion models have showcased various impressive characteristics, \eg, powerful and controllable generative capability, high-quality and diverse output, robust latent space representation, and high flexibility. 
Among them, as illustrated in Figure~\ref{fig:flexibility property}, we argue that the flexibility serves as the core characteristic, where diffusion models generally play the role of a bridge and a connector, being compatible with different upstream input data types (\eg, tabular data, user behavior sequence, multi-modal data, and graph data) and different downstream recommendation models (\eg, general recommenders, sequential recommenders, multi-modal recommenders, and graph-enhanced recommenders). 
This is attributed to the fact that diffusion models are actually a general self-supervised learning paradigm with the fundamental design principle of gradually transforming noise into structured data through an iterative denoising process, which makes the diffusion model a versatile plug-in toolkit for recommender systems. 

\subsection{Diffusion as Recommendation Model}
\label{sec:diffusion as recommender}

The recommendation model serves as the core component of the recommender system pipeline to select or rank the top-relevant items to satisfy users’ information needs based on the outcome of the previous data engineering \& encoding stage. 
When adapting diffusion models as recommendation models, the input for this stage can be structured data (\eg, user behavior sequence), neural embeddings from other encoders, or a combination of both, depending on the architecture design of the diffusion-based recommender. 
In this section, as shown in Figure~\ref{fig:taxonomy}, we mainly focus on the input formats and task formulations of diffusion-based recommendation models, and thereby classify related research works into three categories: (1) collaborative recommendation, (2) context-aware recommendation, and (3) other applications. 

In general, \textit{collaborative recommendation}, also known as collaborative filtering, is the basic recommendation task of modeling user preference primarily based on the user-item interaction records (\ie, user-item co-occurrence matrix). 
It is simple, straightforward, yet effective to capture the overall user preferences via the user-item interactions.
Built upon the basic collaborative signals, \textit{context-aware recommendation} further considers the context information to enable more dynamic and more accurate user interest modelings. 
In this paper, the definition scope of context information is relatively broad, including but not limited to circumstances (\eg, time, location, and device), user temporal dynamic (\eg, behavior sequences instead of a set of records), and item attributes (\eg, textual descriptions and thumbnails). 
Lastly, \textit{other applications} is referred to the tasks that are closely related to the recommendation scenarios but somehow differ from the aforementioned two categories in problem formulations and solution paradigms, such as cross-domain recommendation, learning to rank, and computational advertising. 

Before elaborating on the details of this section, we would like to further clarify the key difference between ``diffusion as recommendation model'' and ``diffusion for data engineering \& encoding''. 
As discussed in Section~\ref{sec:discussion of diffusion for data engineering and encoding}, the methods in the category of ``diffusion for data engineering \& encoding'' focus on producing enhanced structured data or neural embeddings for other downstream recommendation models. 
However, in this section, the diffusion models are directly trained as the recommendation models to generate the user preference distribution over the item space, or produce latent user interest representations for item matching.

\subsubsection{Diffusion for Collaborative Recommendation} 

When adapted for collaborative recommendation, diffusion models generally explore the user preference patterns based on the user behavior history (\ie, user-item interaction matrix)~\cite{walker2022recommendation,wang2023diffusion,benedict2023recfusion,hou2024collaborative,choi2023blurring,jiangzhou2024dgrm,zhu2024graph}. 
They apply the diffusion-then-denoising process to the user interaction records to uncover the potential positive items that users might be interested in. 
According to the evolving modeling paradigms, we roughly categorize the research of this line into three types, and introduce their representative works as follows.

As an earlier attempt, BSPM~\cite{choi2023blurring} takes the \textit{whole user-item co-occurrence matrix} as the exact one input, and carefully designs a perturbation-recovery paradigm, where the interaction matrix is first blurred (perturbed) and then sharpened (recovered) to derive unknown user-item interactions for recommendation. 
However, BSPM differs from classical diffusion model paradigms (\eg, DDPM~\cite{ho2020denoising} or SGM~\cite{song2020score}) in the following two key aspects. 
(1) The blurring and sharpening operations of BSPM are non-parametric methods without training neural networks or learning embedding vectors, while traditional diffusion models have to learn the parametric transformation functions. 
(2) While traditional diffusion models are trained based on a dataset with a large number of images, the input for BSPM in the field of recommender systems is only one user-item interaction matrix. Therefore, the entire blurring-sharpening process can be described by deterministic ordinary differential equations (ODEs). 
However, directly operating over the entire interaction matrix is costly in terms of both memory usage and computational resources, and is therefore non-scalable to scenarios with large-scale users or items (\ie, a giant but sparse interaction matrix).

Instead of manipulating over the entire interaction matrix, other works propose to employ the diffusion-denoising process over the \textit{single user-level interaction records} with parameterized learnable functions~\cite{walker2022recommendation,wang2023diffusion,benedict2023recfusion,jiangzhou2024dgrm,zhu2024graph}. 
DiffRec~\cite{wang2023diffusion} takes as input each user's binary interaction vector, \ie, $\mathbf{x}_0=\{0,1\}^{|\mathcal{I}|}$ with the $i$-th binary element implies whether the user has interacted with item $i$ or not, and adopts the classical DDPM framework to enable the generative recommendation paradigm. 
Two improvements over DDPM are proposed to ensure better personalized recommendations. 
(1) The authors reduce the noise scales and forward diffusion steps to control the corruption over user interaction records. 
(2) DiffRec is optimized by directly predicting the target interaction history $\mathbf{x}_0$ instead of the noise $\epsilon$ to be eliminated at each step, which is more intuitive and empirically training-stable. 
Moreover, DiffRec is further extended into two versions. 
(1) L-DiffRec compresses the input vector via item clustering and thereby enables latent diffusion to reduce the resource costs for large-scale item prediction. 
(2) T-DiffRec incorporates the temporal information by applying positional reweighting to the input interaction vector $\mathbf{x}_0$ to further capture the temporal dynamics of user preferences. 
It is worth noting involving the temporal information actually leads to the field of sequence recommendation (\ie, context-aware recommendation in Section~\ref{sec:context-aware rs}), but the major designs and contributions of DiffRec~\cite{wang2023diffusion} are still within the scope of collaborative recommendation.
Other works of this type generally follow such a setting. 
For instance, CODIGEM~\cite{walker2022recommendation} utilizes multiple autoencoders (AEs) to model the reverse denoising generation yet only leverages the first AE for interaction prediction during the inference phase.
\citet{benedict2023recfusion} propose 1D binomial diffusion to explicitly model the binary user-level interaction vectors with a Bernoulli process.
GiffCF~\cite{zhu2024graph} converts the user-level interaction vector into an item-item similarity graph, and employs a smoothed graph-based diffusion-denoising process using the heat equations, which leverages the advantages of both diffusion models and graph signal processing.

Taking one step further, more recent works~\cite{hou2024collaborative} start to integrate the \textit{high-order multi-hop neighbor information} of the user-item interaction bipartite graph into diffusion-based collaborative recommendation. 
CF-Diff~\cite{hou2024collaborative} uses a cross-attention multi-hop autoencoder to harness the multi-hop connectivity information from the target user during the reverse denoising generation, while making the forward diffusion process remain the same as previous diffusion-based collaborative recommenders like~\cite{wang2023diffusion}. 
This does not only enrich the collaborative signals for denoising generation of users' potential preferences, but also preserves the model complexity at a manageable level.
Furthermore, the authors provide theoretical analysis to prove the computational tractability and scalability of the proposed CF-Diff.

\subsubsection{Diffusion for Context-aware Recommendation}
\label{sec:context-aware rs}

The context-aware recommendation shares the same ultimate goal with the collaborative recommendation, \ie, precisely estimating the user preference towards a certain target item, but differs in the accessible input data. 
While the collaborative recommendation merely utilizes the user-item interaction matrix as input, the context-aware recommendation further takes into account the context information for more relevant and personalized recommendations, \eg,  situational information like time and weather, item attributes like titles and categories, and user profiles like behavior sequences. 
In the following, we discuss the related research works based on different context information that the diffusion-based recommender intends to use.

First and foremost, a number of works aim to involve \textit{temporal/sequential information} for diffusion-based context-aware recommendation models~\cite{yang2024generate,li2023diffurec,du2023sequential,wang2024conditional,niu2024diffusion,wang2024leadrec,li2024dimerec}. 
These studies intend to conduct conditional diffusion-based generative recommendation by taking the representations of user behavior sequences as the guidance for reverse denoising process. 
For example, DreamRec~\cite{yang2024generate} uses a Transformer~\cite{vaswani2017attention} encoder to create guidance representations from user behavior sequences and employs a diffusion model to explore the underlying distribution of item space, generating an oracle next-item embedding that aligns with user preferences without the need for negative sampling. 
As a follow-up work, DiffRIS~\cite{niu2024diffusion} improves the guidance representations from behavior sequences by explicitly modeling the long- and short-term user interests via delicately designed multi-scale CNN and residual LSTM modules.
DimeRec~\cite{li2024dimerec}, on the other hand, strives to optimize the diffusion-denoising process.
To be specific, for each training sample, the model introduces noise to the target item embedding using geodesic random walk~\cite{de2022riemannian} on a spherical space, ensuring isotropic and small noise insertion that approximates Gaussian distribution in the tangent space. 
The reverse denoising process is conducted under the guidance of user sequence representation, and is jointly optimized with the objective of traditional sequential recommender in a multi-task manner. 
Moreover, instead of integrating the sequential behaviors into one compact representation beforehand for guidance, other works~\cite{du2023sequential,li2023diffurec,wang2024conditional} attempt to directly incorporate a sequence of historical item embeddings during the reverse generation process, resulting in implicit conditions on the sequential information.

In addition to the user behavior sequences, researchers also investigate \textit{other diverse context information} for diffusion-based recommenders, such as spatial-temporal information~\cite{qin2023diffusion,long2024diffusion,wang2024dsdrec}, multi-modal knowledge~\cite{yu2023ld4mrec,wang2024dsdrec}, and social networking~\cite{li2024recdiff,he2024balancing}. 
For example, Diff-POI~\cite{qin2023diffusion} concentrates on the point-of-interest (POI) recommendation, and incorporates two specially designed graph encoding modules to encode a user's visiting sequential and spatial characteristics, followed by a diffusion-based sampling strategy to explore the user's spatial visiting trends. 
It uses the diffusion process and its reverse form to sample from the posterior distribution and optimizes the corresponding score-based generative model. 
\citet{long2024diffusion} further propose DCPR~\cite{long2024diffusion} to optimize the on-device POI recommendation by introducing a three-level cloud-edge-device architecture with slightly different diffusion training processes at each level.
The overall framework starts with a global diffusion-based recommender trained with category-level movement patterns on the cloud server, and then customizes for each region on edge servers by considering local POI sequences, and finally finetunes the model on individual devices using personal data.
In this way, DCPR is able to provide region-specific and personalized POI suggestions while reducing computational burdens on user devices. 
LD4MRec~\cite{yu2023ld4mrec} focuses on the multimedia recommendation, and generates the user preference towards the whole item space through the denoising process under the joint guidance of collaborative signals and items' multi-modal knowledge.
The authors also simplify the reverse generation process to enable one-step inference instead of multi-step inference, which greatly reduces the computational complexity.
\citet{li2024recdiff} propose RecDiff for the social recommendation, and design a simple yet effective latent diffusion paradigm to mitigate the noisy effect in the compressed a dense representation space obtained from the user social networks. 
The multi-step noise diffusion and removal is optimized in a downstream task-aware manner, thereby leading to exceptional capabilities in handling the diverse noisy effects of user social contexts.

\subsubsection{Diffusion for Other Applications}

While collaborative and context-aware diffusion-based recommendation models focus on precisely estimating the user preference towards a target item given the in-domain recommendation data (\ie, interaction records and contextual information), diffusion models are also adapted as the core generative functions for other online application tasks.
These tasks are closely related to the recommendation scenarios but somehow
differ from the aforementioned two categories (\ie, collaborative or context-aware recommendation) in problem formulations and solution paradigms.
Hence, we generally categorize them as \textit{other applications} and briefly introduce these research works as follows. 

\citet{lin2024discrete} propose a novel discrete conditional diffusion reranking (DCDR) framework for the \textit{reranking} stage in recommender systems, where the model has to generate a reranked item list. 
DCDR extends traditional diffusion models by introducing a discrete yet tractable forward process with step-wise noise addition through operations at both permutation and token levels. 
It also includes a conditional reverse process that generates item sequences based on the expected user feedback. 
More importantly, DCDR has been deployed in real-world online recommender, where the authors design several inference strategies to satisfy the strict requirements of efficiency and robustness for online applications (\eg, beam search and early stopping). 

DiffBid~\cite{guo2024aigb} leverages diffusion models for \textit{automatic bid generation} in online advertising scenarios, which is called AI-generated bidding (AIGB) in the paper.
The authors address the limitations of traditional reinforcement learning (RL) based auto-bidding methods (\eg, instability in dynamic bidding environments) by introducing a conditional diffusion model to capture the correlation between advertising returns and the entire bidding trajectory. 
To be specific, DiffBid operates by introducing a conditional diffusion process that gradually adds noise to a bidding trajectory in a forward process, transitioning it towards a standard Gaussian distribution, and then progressively denoising it in a reverse process to reconstruct an optimal bidding trajectory. 
The reverse process employs a parameterized neural network conditioned on expected returns and temporal contexts to guide the auto-bidding generation. 
By framing auto-bidding as a diffusion-denoising process, DiffBid can effectively handle the randomness and sparsity inherent in online advertising environments, leading to more stable and efficient bidding strategies. 
It has been deployed on an online advertising platform and has gained significant improvements through the online A/B test.

DiffCDR~\cite{xuan2024diffusion} utilizes diffusion models for \textit{cross-domain recommendation}, aiming to effectively transfer the knowledge from an auxiliary domain to a target domain, especially when dealing with cold-start users who have limited or even no interaction history in the target domain. 
Specifically, DiffCDR generates user embeddings in the target domain by reversing the diffusion process, conditioned on the user's embeddings from the source domain. 
In this way, equipped with powerful generative capabilities to model the underlying data distribution, the diffusion model acts as the core knowledge mapping module to transfer knowledge from source domain to target domain. 
Additionally, the authors also design an alignment loss and a label-data-aware task-oriented loss to further stabilize the training procedure and improve the final cross-domain recommendation performance.

DMSR~\cite{tomasi2024diffusion} attempts to optimize the \textit{slate recommendation} with the help of diffusion models. 
Different from traditional recommendation models that usually estimate the user preference towards each individual item, the slate recommendation aims to optimize the entire collection of items (\ie, a slate/group/bundle) presented to the user at once. 
Hence, the model should comprehensively consider the in-slate mutual effects (\eg, diversity) and overall utilities, thereby suffering from the complex combinatorial choice space. 
To this end, diffusion models turn out to be a promising solution to directly generate a set of latent vectors to construct the slate via item mapping. 
DMSR adopts the diffusion transformer architecture~\cite{Peebles2022DiT}, and conducts the reverse denoising process over the corrupted in-slate item representations under the guidance of contextual information like user preferences or textual queries. 
The denoised latent vectors are then decoded back into the discrete item space, resulting in a slate recommendation that maximizes user satisfaction by balancing the relevance and diversity.

\begin{figure}[t]
    \centering
    \includegraphics[width=0.99\textwidth]{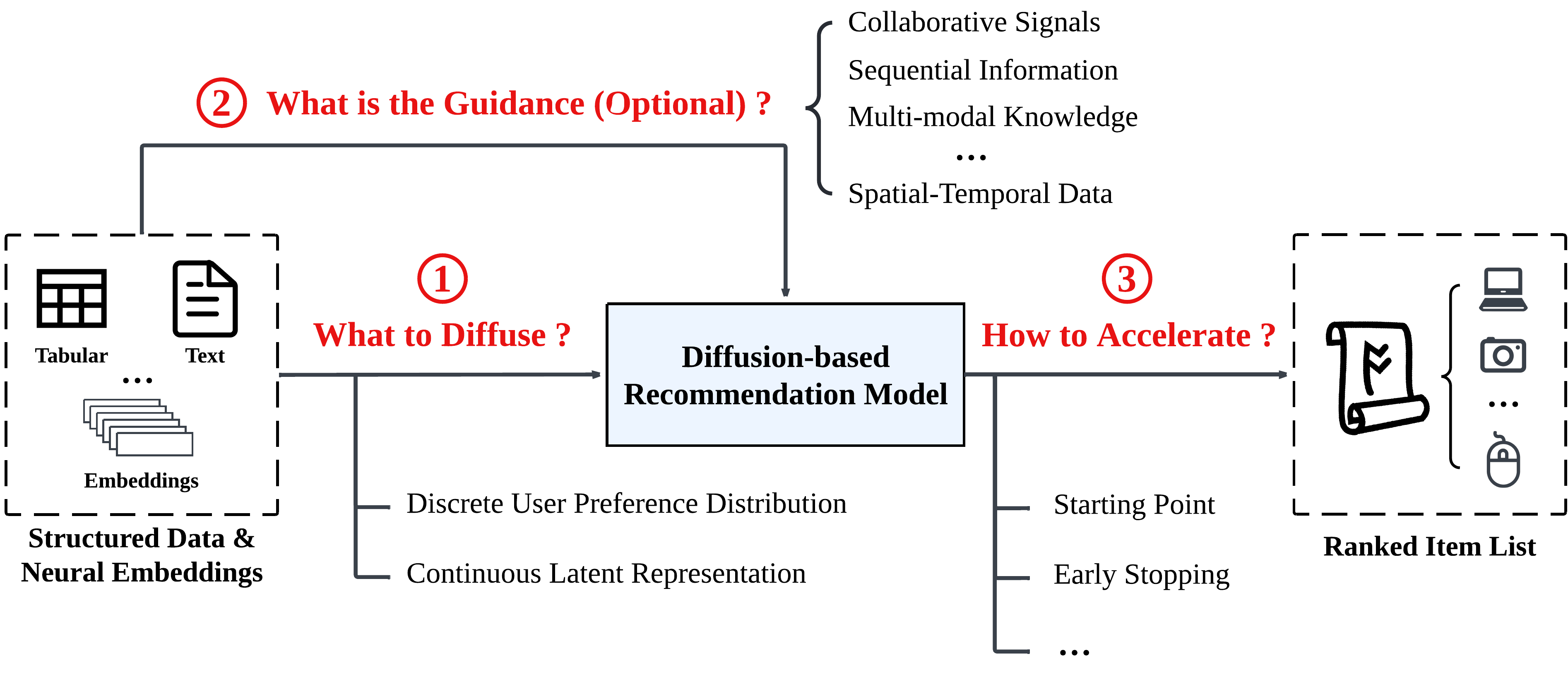}
    \caption{
    The illustration of three key perspectives when adapting diffusion models as recommenders: (1) what to diffuse, (2)what is the guidance (optional), and (3) how to accelerate.
    }
    \label{fig:diff as rec discussion}
\end{figure}

\subsubsection{Discussion}

Due to the exceptional generative capabilities and flexibility with different backbones and downstream tasks, diffusion-based recommendation models have turned out one of the the state-of-the-art generative recommendation paradigms by learning the underlying distribution of user preferences and item characteristics. 
From the aforementioned discussion, we can clearly observe the potential advantages through the adaptation of diffusion models as recommenders, including but not limited to capturing the complex and non-linear relationships between users and items and therefore generating diverse and novel recommendations.
In this section, as shown in Figure~\ref{fig:diff as rec discussion}, we would like to further discuss this line of research from the following three key perspectives when constructing the diffusion-based recommendation models: (1) what to diffuse, (2) what is the guidance, and (3) how to accelerate.

\textbf{What to diffuse?} 
As shown in Figure~\ref{fig:diff as rec discussion}, there are generally two types of inputs for the diffusion process: discrete user preference distribution and continuous latent representations. 
The first choice is to diffuse the interaction matrix or the probabilistic distribution of user preferences over the entire item space~\cite{benedict2023recfusion,wang2023diffusion}, \eg, collaborative signals $\mathbf{x}_0=\{0,1\}^{|\mathcal{I}|}$. 
Although such kind of methods is simple and straightforward to directly denoise and generate the target user preference distribution towards potentially unobserved interactions, it is not scalable when facing a large volume of items due to the tremendously large input space and extensive resource requirements.
Moreover, other contextual information like multi-modal knowledge can only be involved in the guidance part, which limits the model design. 
As a result, the second choice proves a more coherent, scalable and commonly adopted solution to conduct the diffusion-denoising process in the continuous latent space~\cite{yu2023ld4mrec,wang2024dsdrec,li2024recdiff,he2024balancing}. 
In this way, the diffusion-based recommenders are able to first integrate multi-source knowledge (\eg, spatial-temporal data and social networks in addition to collaborative signals) into a unified representation via specially designed encoders, and employ the diffusion model to learn the underlying latent distributions, which is more compatible with various tasks like POI recommendation or social recommendation. 
The denoised latent representations, usually serving as the user preference vectors or target item representations, are then used to retrieve the top-$K$ relevant items through vector matching for the final ranked item list.

\textbf{What is the guidance (optional)?} 
As depicted in Figure~\ref{fig:diff as rec discussion}, the guidance serves as the conditioning information to tailor the reverse denoising process for either discrete user preference distributions or continuous latent representations. 
It plays a crucial role in ensuring that the generated recommendations are not only high-quality but also relevant and personalized for each user. 
Although optional, the guidance mechanism has been widely adopted in diffusion-based recommendation models, especially for context-aware recommendation where various context information can be fused into the conditioning signals, \eg, sequential information~\cite{yang2024generate,li2024dimerec}, multi-modal knowledge~\cite{yu2023ld4mrec,wang2024dsdrec}, and spatial-temporal data~\cite{qin2023diffusion,long2024diffusion,wang2024dsdrec}.

\textbf{How to accelerate?} 
By determining the aforementioned two key factors (\ie, contents to be diffused and the optional guidance), one can already construct the overall architecture of diffusion-based recommendation models. 
However, the inference efficiency is another important topic when directly adapting diffusion models as recommenders, since online recommender systems are usually real-time services and extremely time-sensitive, where the user request should be responded within around tens of milliseconds~\cite{lin2024can,lin2023map}. 
Diffusion models can be computationally expensive due to the multi-step denoising process, especially when dealing with large-scale recommender systems. 
Several strategies have been proposed to balance the trade-off between recommendation quality and computational efficiency. 
For example, \citet{lin2024discrete} and \citet{wang2023diffusion} propose to start the denoising inference from a meaningful input (\eg, outputs from the previous stages or partially corrupted user preference representations) instead of the pure Gaussian noise in traditional diffusion models, which does not only avoid totally eliminating the important personalized information, but also accelerate the inference phase with much fewer denoising steps. 
Other works also attempt to speed up the denoising process by adopting the early-stopping strategy~\cite{lin2024discrete} or one-step inference~\cite{yu2023ld4mrec}.
Despite these preliminary explorations, there is still a lack of systematic studies on the general acceleration strategies for the inference of diffusion-based recommendation models, formulating a promising direction for future research over which we would cast further discussions in Section~\ref{sec:challenge}.

\subsection{Diffusion for Content Presentation}
\label{sec:diffusion for content presentation}

The two stages discussed above (\ie, data engineering \& encoding, and recommendation models) concentrate on selecting and arranging the optimal ranked item list for recommendation. 
In this section, the phase of content presentation focuses on customizing and seeking the best presentation strategy of recommended items for different users (\eg, individualized titles or thumbnails). 
This is non-trivial to not only consider the user's potential preferences based the historical and contextual information but also involve the visual appearance correlation and coherence. 
While earlier attempts for content presentation heavily rely on the manual design and pre-defined page arrangement, the emerging of top-tier generative models, especially diffusion models like Stable Diffusion~\cite{rombach2022high}, points out a promising direction to automate and even personalize the content generation for the item display. 
According to the different types of contents to be generated by diffusion models, the existing research works in this line roughly exhibits three progressive development stages: (1) fashion generation, (2) ad creative generation, and (3) general content generation. 

Before we elaborate on the details, we would like to clarify the key difference between diffusion for content presentation in this section and diffusion for data augmentation in Section~\ref{sec:diffusion for data augmentation}, since both of them involves using diffusion models to generate synthetic images or other contents. 
The core difference lies in the usage of the generated materials. 
When leveraging diffusion models for data augmentation, we want the synthesized content to be the model input and affect the downstream recommenders for various purposes, \eg, privacy preserving, adversarial attack, or recommendation enhancement. 
Instead, when employing diffusion models for content presentation, we would assume that the recommended items have already been chosen and ranked for different users, and mainly focus on further affecting and promoting the user satisfaction by presenting the automatically generated content to the target user. 
Moreover, since this line of research is under-explored and possess only a few related works~\cite{czapp2024dynamic,yang2024new,shilova2023adbooster,xu2024diffusion,mukande2024mmcrec,shen2024pmg,wang2023generative}, in the following, we would also introduce and discuss the previous works that leveraging other generative models (\eg, adversarial generative networks, or language models) for content presentation, which facilitates the elucidation of the development path within this research area. 


\subsubsection{Fashion Generation}

Fashion recommendation is a special domain that emphasizes on the complementary relationships and visual presentations for recommending a personalized fashion outfit.
Hence, it is straightforward and natural to introduce the generative models to synthesize real-looking fashion clothes. 
This would inspire the aesthetic appeal and curiosity of both costumers and designers, and motivates them to explore the space of potential fashion styles, finally leading to improved recommendation performance. 


Before diffusion models emerge as the prominent methods among generative models, earlier works generally rely on generative adversarial network (GAN) based frameworks to generate the fashion outfits for recommendation~\cite{deldjoo2021survey}. 
For instance, DVBPR~\cite{kang2017visually} leverages GANs' visual generative capabilities to synthesize clothing images based on user preferences for fashion recommendation. 
It trains a GAN-based framework with an integrated user preference maximization objective, and is able to generate realistic and plausible fashion images that better align with user preferences compared to the original manually designed clothing materials. 
\citet{shih2018compatibility} designs a compatibility learning framework to allow the users to visually explore candidate compatible prototypes (\eg, clothing collocation for a white T-shirt and blue jeans). It takes as input a prototype representation encoded from a query image of clothes, and uses metric-regularized conditional GAN (MrCGAN) to produce a synthesized image of a complementary item across various categories. 
Other works also focus on the fashion complementary generation problem with GAN frameworks by introducing Bayesian personalized ranking~\cite{yang2018recommendation} and randomized label flipping~\cite{kumar2019c}.

While GAN-based models generally suffer from the training instability problem, diffusion models turn out a more promising solution for fashion generative recommendation. 
\citet{xu2024diffusion} propose to develop generative outfit recommendation based on diffusion models. 
The proposed DiFashion framework can not only complete the fashion complementary item generation task, but also produce personalized outfit images from scratch according to the user preference. 
The model finetunes the latest Stable Diffusion SD-v2~\cite{rombach2022high} to ensure the high fidelity, compatibility, and personalization of generated fashion images.
Besides, three types of conditions (\ie, category prompts, mutual outfit conditions, and historical conditions) are introduced to jointly guide the denoising generation process, ensuring the quality, internal consistency, and alignment with user preferences, respectively. 


\subsubsection{Ad Creative Generation}

Compared with fashion outfits, ad creative generation covers a broader range of visual contents, including but not limited to apps, games, cosmetics, and other promotional images. 
Previously, the creatives of ad campaigns are designed with great care and manual efforts.
However, this human-designer-centered approach are non-scalable and suboptimal, because the designer often struggles to fully consider the global preferences of targeted user groups, and also fails to adapt to different online advertising scenarios. 
A common solution is to show the original product image with additional design elements adaptively generated by some neural models, \eg, language models to customize the headlines and captions~\cite{mita2023camera,kanungo2021ad}. 
Moreover, since the ad creative aims to promote the sale or click rate of a specific product, there are usually more constraints for automatic ad creative generation. 
For example, we have to maintain the original target product within the generated or edited creative image. 
Besides, the seller and designer might have additional guidelines and instructions when employing artificial intelligence generated content (AIGC) for creative modification or generation, \eg, the overall visual styles and color themes should be coherent and consistent. 
To this end, diffusion models are becoming increasingly popular to facilitate the ad creative generation with inpainting/outpainting mode~\cite{lugmayr2022repaint} and text-to-image controllable generation~\cite{abdollahpouri2017controlling}. 


\citet{shilova2023adbooster} propose the generative creative optimization task, and utilize user interest signals to personalize the ad creative generation with the outpainting mode of stable diffusion~\cite{rombach2022high}. 
The proposed AdBooster first masks out the background of the original product image (\eg, a model with certain fashion outfits), and then finetunes a stable diffusion model to outpaint the masked background with textual prompt guidance based on the user query and contextual information. 

\citet{yang2024new} also follow such a background generation paradigm, and further refine the solution by introducing a new automated creative generation for click-through rate (CTR) optimization pipeline (CG4CTR). 
Specifically, the authors apply the inpainting mode of stable diffusion to generate the background images while keeping the main product details unchanged. 
The stable diffusion model is finetuned with two assistant models, \ie, prompt model and reward model. 
The prompt model is designed to generate personalized textual prompt guidance for different user groups to enhance the diversity and quality of stable diffusion generation.
The reward model is a pretrained model to output the CTR score for each generated creative image, considering multi-modal features of both images and texts. 
It plays a critical role to select the best creatives according to the estimated CTR scores for training and online display. 
During training, the stable diffusion model and prompt model are iteratively updated in turn based on the reward signals from the frozen reward model. 
The proposed CG4CTR framework has been deployed and validated on a large-scale e-commercial platform. 

More recently, \citet{czapp2024dynamic} employs diffusion models for the creation of eye-catching personalized product images to increase user engagement with recommendations in online retargeting campaigns for real-world industrial applications.
Although still adopting the basic fill-in-the-background paradigm for ad creative generation, the authors improve the overall pipeline from the following aspects. 
Firstly, the position and size of the masked product object would be further adjusted to the center of the image according to the placement for emphasis. 
Secondly, an optional edge detection is introduced to further reinforce the contours with the background mask. 
Finally, the authors use LinUCB~\cite{li2010contextual,chu2011contextual} contextual multi-armed bandit algorithm to select the prompt with the highest predicted CTR from a pre-defined pool of prompts associated with the product category in the given context (defined by user, item, and location features), which is designed to balance the generative personalization and online latency constraint.

\subsubsection{General Content Generation}

The aforementioned fashion and ad creative generation primarily focus on generating or editing materials of visual modality for items with open-sourced stable diffusion models~\cite{rombach2022high}. 
Encouraged by the booming of generative multi-modal foundation models like Sora~\cite{sora} and Transfusion~\cite{zhou2024transfusion}, we are able to further push the boundaries of what is possible in diffusion-based AIGC, with capabilities of generating much more generalized contents of various hybrid modalities, \eg, audio, vision, and video. 

The exploratory research works~\cite{mukande2024mmcrec,shen2024pmg,wang2023generative} generally follow the similar framework, where the diffusion-based content generator is guided by the conditions extracted by a hybrid instructor module. 
The content to be generated by diffusion models is no longer limited to images of clothing or ad creatives, but has already extended to a broader range of modalities. 
The hybrid instructor module usually involves large language models (LLMs) to deepen the understanding of user intents assisted by other traditional models like sequence models and graph-enhanced models. 
Built upon this basic setting, \citet{wang2023generative} propose GereRec to further refine the formulation and deepen the integration of the diffusion-based AIGC module with the overall recommendation pipeline. 
GereRec not only generates the personalized content, but also supports a variety of operations to further adjust the item content to maximize the user-oriented utilities, \eg, thumbnail generation and selection, caption customization, and domain-specific fidelity checks. 
Although the research in this line is still preliminary and not well validated on large-scale platforms, we believe this is a direction worth delving into, which could potentially lead to the next-generation generative recommender paradigms.

\begin{figure}[t]
    \centering
    \includegraphics[width=0.99\textwidth]{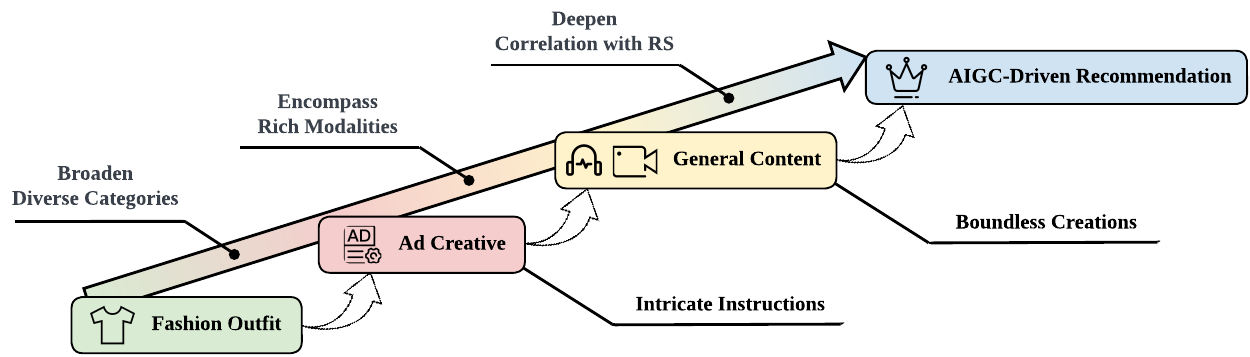}
    \caption{
    The illustration of the development trend for adapting diffusion models for content presentation. 
    }
    \label{fig:content presentation discussion}
\end{figure}

\subsubsection{Discussion}
\label{sec:discussion of diffusion for content presentation}

Generally speaking, adapting diffusion models for content presentation highlights an important evolution in recommender systems. It is no longer limited to selecting and ranking candidate items, but further delves into the user interaction phase and thereby dynamically customizes the content displays of the same items for different users.
As shown in Figure~\ref{fig:content presentation discussion}, we could observe a clear development trend for adapting diffusion models for content presentation from the perspective of generated contents. 
Earlier studies begin with generating images primarily related to fashion clothing.
Then, researchers expand the generative scope and encompass a broader range of ad creatives for online advertising. 
In this circumstance, diffusion models have to generate or edit the image of a certain product under various pre-defined guidelines and intricate instructions, \eg, maintaining the details of the original product, and preserving the coherent and consistent background rendering with the commands from the users or designers. 
Next, with the emergence of multi-modal foundation models, it evolves beyond the visual modality, and starts to incorporate more generalized and diverse multi-modal data generation, including but not limited to video and audio. 
In this way, although it imposes much higher demands on the capabilities of generative models, the content generation process itself has become relatively less constrained, allowing for greater creation freedom. 
This, in turn, is more conducive to dynamically synthesizing content that aligns with users' personalized preferences. 
Finally, we suggest that this progression points toward the ultimate goal of AIGC-driven recommender systems when the correlation and cooperation between the diffusion-based content generation is further deepened with the overall recommendation pipeline. 
This signals not just the technological progress, but also a profound shift in how humans interact with AI and digital creativity, leading to a new paradigm in the digital content ecosystem.
In such a rapidly emerging field of research, there are naturally still many challenges and issues that urgently need to be addressed, which will be discussed in Section~\ref{sec:challenge}.


\section{Future Directions and Challenges}
\label{sec:challenge}

In this section, we cast detailed discussions on future directions on the research of adapting diffusion models to recommender systems.
We would highlight the key challenges, and further discuss the preliminary efforts done by existing works, as well as other possible solutions. 

\subsection{Efficiency and Scalability}

Diffusion models excel at learning the underlying distribution of the training data, which necessitates the distribution sampling during their denoising generation process. 
One of the major drawbacks of the diffusion models is that their sampling process is fairly inefficient, thereby resulting in slow data generation. 
This inefficiency issue stems from their reliance on a long Markov chain of diffusion steps for sample generation, which is both time-consuming and computationally expensive~\cite{ulhaq2022efficient}. 
However, as typical real-time services, recommender systems are extremely time-sensitive and resource-constraint in terms of the large-scale users/items, which casts a significant challenge about the efficiency and scalability of diffusion-based recommender systems.
Moreover, when adapting diffusion models to different stages of the recommendation pipeline as discussed in Section~\ref{sec:taxonomy}, we can observe different forms of such a challenge of efficiency.
\begin{itemize}
    \item When adapting diffusion models for data engineering \& encoding, the inefficiency issue mainly affects the \textit{training phase} of recommender systems, since the output of diffusion models (either augmented data or enhanced representations) can be pre-computed and pre-cached to avoid the influence towards the online inference stage. 
    However, the slow generation can limit the training efficiency of downstream recommendation models, which typically requires both the large volumes of training data (million- or even billion-level) and the update frequency (from day-level to hour-level or even minute-level). 
    Existing works~\cite{wang2024diff,liu2023diffusion} generally adopt the asynchronous update strategy to separate the training of diffusion models and downstream recommenders. 
    In this way, we can reduce the training data volume and relax the update \& generation frequency for diffusion models, while maintaining full training data and high update frequency for downstream recommenders. 
    \item When adapting diffusion models as recommendation models, we have to take into account the \textit{inference efficiency} problem, since the diffusion-based recommenders are directly deployed for real-time online services, where the user request should be responded to within around tens of milliseconds. 
    In this context, preliminary works have been conducted to reduce the inference latency with few-step~\cite{wang2023diffusion} or even one-step sampling process~\cite{yu2023ld4mrec}, and possibly combined with an early-stopping strategy~\cite{lin2024discrete}.
    \item When adapting diffusion models for content presentation, the \textit{storage overhead} should be further considered.   
    As discussed in Section~\ref{sec:discussion of diffusion for content presentation}, given the fact that we are currently unable to support high-resolution real-time online content rendering for each user request (\ie, inference inefficiency), we have to pre-compute and pre-cached the personalized content beforehand. 
    The operations of pre-storing individualized contents could then meet the storage overhead, \ie, we have to expand the space complexity from $O(M)$ for $M$ items to $O(MN)$ for $M\times N$ user-item pairs. 
    While generating personalized item contents for each single user is impractical due to such a storage constraint, existing works~\cite{yang2024new,wang2023generative} intend to perform user clustering and pre-cache the group-wise personalized item contents.
\end{itemize}

Based on the specific efficiency challenges and preliminary solutions at different recommendation stages discussed above, future directions can explore lighter and faster versions of diffusion model architectures with the various techniques like model compression~\cite{wang2024patch} and knowledge distillation~\cite{luhman2021knowledge}. 
Besides, further efforts on efficient sampling strategies (\eg, parallel computing~\cite{li2024distrifusion}, denoising scheduler~\cite{ma2024deepcache}, and retrieval strategy~\cite{chen2022re,rombach2022text}) are also of great importance to address the inefficiency issue of diffusion-based recommender systems.

\subsection{Integration with Large Language Models}

Although diffusion models excel at capturing complex data distributions and producing diverse outputs across various modalities (\eg, text, images, as well as latent representations), most diffusion-based recommender systems still operate in a closed system, learning only from narrowly defined in-domain data~\cite{xi2023towards,lin2024rella}. 
Such closed-system approaches restrict the adaptability of the systems and can lead to suboptimal recommendation results. 
To this end, the integration of large language models (LLMs) opens up the opportunity to access the external open-world data, allowing recommender systems to acquire knowledge beyond their pre-defined boundaries. 
This helps establish a more flexible and context-aware recommendation framework that better aligns with user needs. 

To be specific, LLMs are skilled at interpreting user profiles, queries, and historical interactions by analyzing natural languages, detecting behavior patterns, and discerning subtle semantic cues~\cite{lin2024can}. 
This integration is particularly important in scenarios where the user's preferences are complex and dynamic, requiring both a deep understanding of user intent and the ability to generate a wide range of diverse content.
In this way, LLMs serve as a bridge between user inputs and the generative capabilities of diffusion models. 
For instance, by understanding the nuances of a user's intent, LLMs can convert natural language prompts into meaningful guidance for diffusion-based recommendation models, ensuring that the generated recommendations (\eg, ranked item list) or personalized contents (\eg, thumbnails) are not only diverse but also contextually tailored for the user preference. 
Apart from continuous-space guidance generation, LLMs are also capable of constructing and reformulating individualized prompts for controllable text-to-image generation~\cite{ma2024exploring} to improve the quality of content personalization for recommendation. 
Moreover, the combination of LLMs and diffusion models points towards a more sophisticated and automated reasoning system, which is able to provide more interactive, personalized, and coherent multi-modal recommendations.

\subsection{Explainability and Interpretability}

Explainability and interpretability are crucial factors that require the personalized recommender systems to not only provide recommendation results, but also further explain and clarify why such items are recommended~\cite{zhang2020explainable}. 
In this way, we can improve the transparency, trustworthiness, persuasiveness, and user satisfaction of recommender systems, and also facilitate system managers to diagnose, debug, and refine the recommendation algorithm. 
Although diffusion models have proven to be a promising solution for recommendation, they generally suffer from the unexplainability due to their black-box generative nature and the stochastic sampling of the diffusion-denoising process. 
While it is pivotal to establish the explainable diffusion-based recommendation, there is only one preliminary work~\cite{guo2023explainable} that explores to enhance the interpretability of diffusion recommenders by training a textual decoder to generate the explanation based on the denoised user representation. 

As for future directions, we suggest that there are two vital types of techniques to improve the explainability of diffusion-based recommender systems, \ie, large language models~\cite{zhao2023survey} and causal learning~\cite{kaddour2022causal}. 
On the one hand, large language models have shown impressive capacities in generating human-like texts for a wide range of tasks. 
Through techniques like prompt learning, LLMs can efficiently adapt to specific recommendation tasks without extensive re-training, and allow for the creation of coherent and contextually relevant justifications that improve the transparency and user satisfaction~\cite{chen2023survey,luo2023unlocking,ma2024exploring}. 
On the other hand, causal learning and counterfactual reasoning could discern and identify the causal relationships or inter-dependencies among variables within the given data, and make counterfactual predictions under different circumstances. 
Hence, incorporating diffusion-based recommendation with causal learning and counterfactual reasoning methodologies can harness the cause-and-effect relationships and counterfactual estimation rather than simple denoising generation, thereby leading to more reliable and interpretable recommendation results~\cite{wu2022opportunity,xu2021learning,tan2021counterfactual}.

\subsection{Digital Copyright and Privacy Preserving}

When adapting diffusion models for content presentation, the multi-modal material generation can cause significant digital copyright challenges, particularly related to data ownership, derivative works, and fair use. 
The diffusion models are often trained on vast datasets that include copyrighted materials, raising the risk of infringement if used without proper permissions. The resulting personalized content may also be considered a derivative work, complicating the issue of who holds ownership—whether it's the original content creator, the model developer, or the end user. 
Furthermore, the blending of various copyrighted data sources in multi-modal outputs makes it difficult to trace and attribute the original creators, potentially violating copyright laws. 
The concept of ``fair use'' is frequently cited as a defense, but the boundaries between transformative AI-generated content and infringement remain ambiguous for diffusion-based recommender systems. 
Besides, the adaptation of diffusion models for either data engineering \& encoding or as recommendation models can also meet the challenge of privacy preserving, where the users' sensitive personal data should be well preserved from leaking. 

To address these challenges, one obvious solution is to source the training data from copyright-free or permissively licensed content, such as public domain works or those under Creative Commons licenses, which can help avoid infringing on the intellectual property rights of creators. 
Moreover, some cutting-edge research works start to investigate the neural automated watermarking~\cite{liang2023mist} or text editing strategies~\cite{somepalli2023understanding} to ensure that the generated content properly attributes the original creators, therefore helping address concerns over authorship.
For instance, \citet{liang2023adversarial} proposes to employ adversarial samples to add imperceptible perturbations to human-created artworks, which can disturb the training of diffusion models. 
In this way, the authors establish a powerful toolkit for human creators to protect their artworks from being used without authorization by diffusion-based AIGC applications.
In the future, we suggest that the hybrid techniques of neural watermarking~\cite{min2024watermark}, adversarial samples~\cite{costa2024deep}, federated learning~\cite{li2021survey} and differential privacy~\cite{ji2014differential} should be vital to mitigate the challenges of digital copyright and privacy preserving in the research field of diffusion-based recommender systems.

\section{Conclusion}
\label{sec:conclusion}

In this survey, we provide a comprehensive review of the research efforts on adapting diffusion models to recommender systems. 
We systematically classify existing research works into three primary categories: (1) diffusion for data engineering \& encoding, which focuses on data augmentation and representation enhancement; (2) diffusion as recommendation models, which employs diffusion models to directly estimate user preferences and rank items; and (3) diffusion for content presentation, which leverages diffusion models to generate personalized content such as fashion and advertisement creatives. 
We also give detailed discussion about the core characteristics of the adapting diffusion models for recommendation, and further identify key challenges and future directions for exploration. 
We hope this survey could serve as a foundational roadmap for researchers and practitioners to advance recommender systems through the innovative application of diffusion models.



\bibliographystyle{ACM-Reference-Format}
\bibliography{sample-base}

\end{document}